\documentclass[journal,twoside,web]{IEEEtran}
\usepackage{cite}
\usepackage{amsmath,amssymb,amsfonts}
\usepackage{algorithmic}
\usepackage{graphicx}
\usepackage{subfig}
\usepackage{multirow}
\usepackage{color}
\usepackage{bbm}
\usepackage{soul}
\usepackage{eurosym}
\usepackage{comment}
\DeclareMathOperator*{\argmax}{argmax}
\usepackage{textcomp}
\def\BibTeX{{\rm B\kern-.05em{\sc i\kern-.025em b}\kern-.08em
    T\kern-.1667em\lower.7ex\hbox{E}\kern-.125emX}}
\begin{document}
\title{A Neural Network based Framework for  Effective  Laparoscopic Video Quality Assessment}
\author{Zohaib Amjad Khan, Azeddine Beghdadi, Mounir Kaaniche, \\ Faouzi Alaya Cheikh, and Osama Gharbi.  \vspace{-1cm} \thanks{ ``This work was carried out within the framework of European ITN project HiPerNav. It has received funding from the European Union’s Horizon 2020 Research and Innovation programme under Marie Skłodowska-Curie grant agreement No. 722068.'' }
\thanks{ Z. A. Khan is with the L2TI, Institut Galil{\'e}e, Universit{\'e} Sorbonne Paris Nord, France (e-mail: zohaibamjad.khan@univ-paris13.fr). }
\thanks{A. Beghdadi is with the L2TI, Institut Galil{\'e}e, Universit{\'e} Sorbonne Paris Nord, France (e-mail: azeddine.beghdadi@univ-paris13.fr).}
\thanks{M. Kaaniche is with the L2TI, Institut Galil{\'e}e, Universit{\'e} Sorbonne Paris Nord, France (e-mail: mounir.kaaniche@univ-paris13.fr).}
\thanks{F. Alaya Cheikh is with the Norwegian Colour and Vision Computing Laboratory, NTNU, Gj{\o}vik, Norway (e-mail: faouzi.cheikh@ntnu.no).}
\thanks{O. Gharbi is with the L2TI, Institut Galil{\'e}e, Universit{\'e} Sorbonne Paris Nord, France (e-mail: osamaabderrahman.gharbi@edu.univ-paris13.fr).}}

\maketitle

\begin{abstract}
Video quality assessment is a challenging problem having a critical significance in the context of medical imaging. For instance, in laparoscopic surgery, the acquired video data suffers from different kinds of distortion that not only hinder surgery performance but also affect the execution of subsequent tasks in surgical navigation and robotic surgeries. For this reason, we propose in this paper neural network-based approaches for distortion classification as well as quality prediction. More precisely, a Residual Network (ResNet) based approach is firstly developed for simultaneous ranking and classification task. Then, this architecture is extended to make it appropriate for the quality prediction task by using an additional Fully Connected Neural Network (FCNN). To train the overall architecture (ResNet and FCNN models), transfer learning and end-to-end learning approaches are investigated. Experimental results, carried out on a new laparoscopic video quality database, have shown the efficiency of the proposed methods compared to recent conventional and deep learning based approaches. 
\end{abstract}

\begin{IEEEkeywords}
Video quality assessment, distortion classification, quality prediction, residual networks, fully connected neural network, end-to-end learning. \vspace{-0.2cm}
\end{IEEEkeywords}

\section{Introduction}
\label{sec:intro}

\begin{figure}[ht]
	\centering
	\includegraphics[width=0.5\textwidth]{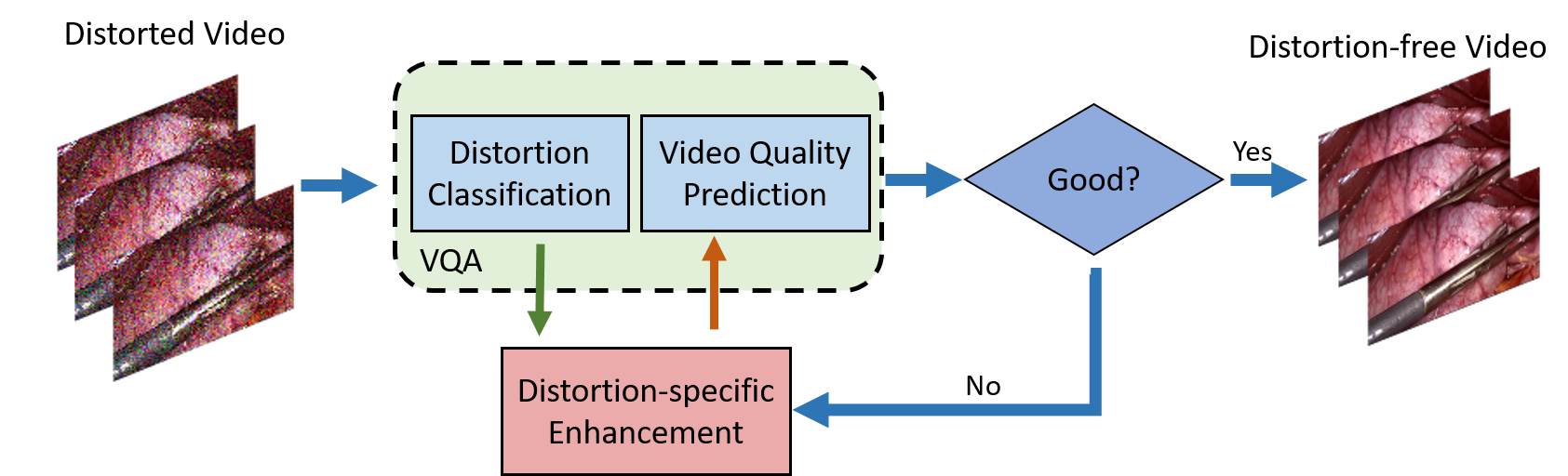}
	\caption{Basic quality monitoring and correction pipeline.}
	\label{fig:quality_monitor}
	\vspace{-0.2cm}
\end{figure}

\IEEEPARstart{M}{onitoring} and assurance of a good video quality is a very critical task in various applications like video streaming, video surveillance, medical procedures and underwater exploration\cite{wang2002image}\cite{wang2011applications}. Any loss of useful information in these videos, resulting from some kinds of video distortion, may not only affect the visual experience but could also cause fatal consequences like in the fields of video surveillance and medical imaging. 
For instance, one of the most common medical procedures, where having a good video quality becomes highly desirable, is the laparoscopic/endoscopic surgery. The latter is a Minimally Invasive Surgery (MIS) which relies on tiny incisions that allow surgeons to enter their surgical instruments inside the body. Among these instruments is the endoscope which is equipped with a camera and is used for the surgical planning and navigation. Hence, a bad video quality can not only cause visual discomfort but can also seriously hamper a surgeon's ability to perform tasks accurately. Such quality may also affect some subsequent processing steps during navigation like registration and segmentation. Moreover, during robotic and image-guided laparoscopic surgeries, there is a need to track instruments in the video. However, the distortions in the captured  video are considered as a main obstacle for achieving accurate track of instruments \cite{sanchez2011laparoscopic} \cite{nwoye2019weakly}. For these reasons, it is imperative to have a quality monitoring and correction system that predicts the video quality, classifies the distortion types in the video and removes them using the appropriate techniques \cite{sdiri2018efficient} \cite{beghdadi2018towards}. Figure \ref{fig:quality_monitor} illustrates the basic idea of such a quality monitoring and correction system.

Generally, Video Quality Assessment (VQA) has been for more than three decades an active field of research \cite{wolf1991objective,hou2020no}. Despite recent progress, VQA remains a challenging problem that seeks for more efficient solutions. VQA requires integrating more high-level semantic visual information in a well explainable learning process to boost the quality prediction model. Existing VQA methods are generally categorized into three classes, namely Full-Reference (FR), Reduced-Reference (RR) and No-Reference (NR), depending on the availability of the reference video. 
 
However, in many real applications, the reference video  is not available and hence, NR metrics \cite{zhu2014no} are the only plausible solutions as they do not rely on such prior information. Before providing an overview of the VQA methods, let us emphasize here that most of the developed works have been devoted to natural video data \cite{tu2020ugc,chikkerur2011objective,gotz2019no, li2019quality, gotz2021konvid} and only few works have focused on medical data, but most often dedicated to coding and transmission problems. \cite{nouri2010subjective, razaak2014study,leveque2017study}. 
\vspace{-0.2cm}
\subsection{Conventional VQA methods}
Research works on VQA have always closely associated it to its image counterpart (i.e Image Quality Assessment (IQA)). A simple and common approach consists first in applying existing IQA methods to each frame and then taking the arithmetic mean of the scores corresponding to all frames to get the final quality score of the video. This is often considered as a kind of a temporal pooling step referred to as average (or arithmetic mean) pooling. In addition, other methods that aim to extend some conventional IQA algorithms for VQA puropose have also been developed. For example, in~\cite{wang2004video}, Wang \textit{et al.} have proposed an extended version of the well known SSIM metric~\cite{wang2004image} to measure the video quality at three levels namely the local region level, the frame level and the sequence level. However, it is not so efficient to address this complex research problem on VQA by simply resorting to simple extension of IQA while neglecting the temporal aspects in the video data and their effect on the global perceptual quality. These spatio-temporal phenomena are closely linked to high-level processes involving learning in the visual cortex. This motivates the use of learning approaches in VQA as well as temporal pooling strategies based on human psycho-visual reasoning like temporal hysteresis \cite{seshadrinathan2011temporal} or fusion-based ensemble pooling~\cite{tu2020comparative}.
To effectively take into account the temporal information in VQA, some metrics based on hand-crafted features exist for natural videos. Indeed, video BLIINDS (V-BLIINDS) \cite{saad2014blind} is a NR VQA metric that uses a combination of motion coherency features and spatial-temporal features based on natural scene statistics. Moreover, MOtion-based Video Integrity Evaluation (MOVIE) \cite{seshadrinathan2009motion} metric and  video intrinsic integrity and distortion evaluation oracle (VIIDEO)~\cite{mittal2015completely} are two other metrics that also use hand-crafted features. MOVIE index is a FR metric that aims to capture the characteristics of Human Visual System (HVS) for video perception by modeling them using separable Gabor filter banks. VIIDEO, on the other hand, is a NR opinion-unaware metric that models statistical naturalness by exploiting space-time statistical regularities.

\vspace{-0.2cm}
\subsection{Neural networks-based VQA methods}
While these hand-crafted features-based VQA methods perform well on the types of distortion and content for which they are designed, they may fail for other kinds of distortion and content. To overcome this drawback, a particular attention has recently been paid to the use of deep neural networks for VQA \cite{hou2020no,liu2018end,ahn2018deep,varga2019no,gotz2019no}. It is important to note here that most of these works only focus on VQA for compressed videos at various severity levels and have not been targeted and tested for other kinds of distortion. For instance, in \cite{liu2018end}, the authors have proposed a single end-to-end optimized  network for compressed videos called V-MEON (Video Multi-task End-to-end Optimized neural Network). Their network is composed of three main blocks used for spatial-temporal feature extraction, codec classification and quality prediction, respectively. 
Besides that, some other works make use of neural networks only for feature extraction. For example, in~\cite{ahn2018deep}, the authors have proposed a hybrid approach, called Deep Blind VQA (DeepBVQA),  whereby they have first extracted spatial features from multiple patches for each frame using a pretrained Convolutional Neural Network (CNN). These features are then fused with hand-crafted temporal sharpness variation feature.
Moreover, Varga has proposed in \cite{varga2019no} a different strategy based on temporal pooling of the frame level deep features. Indeed, for extracting deep features from frames, pretrained CNNs (Inception-V3 and Inception-ResNet-V2) are fine-tuned to classify the quality of the frame into one of the five predefined levels. In a similar approach, the authors in \cite{gotz2019no} have used a pretrained  Inception-ResNet-V2 to extract features from multiple layers of this model. Then, instead of temporal pooling, they have tested two strategies. In the first one, the average feature vector is used as an input of a feed-forward neural network. In the second strategy, they have used deep Long Short Term Memory (LSTM) architecture. However, both these methods\cite{varga2019no} \cite{gotz2019no} give good results for one video quality database and do not perform well on the other tested databases. 
More recently, Hou \textit{et al.} \cite{hou2020no} have proposed a two-stage network for NR-VQA. Thus, they have first used 12 layers of VGG-Net for extracting frame-based features which are then fed into a 3D CNN. The network is pretrained on ImageNet dataset and tested only on LIVE, CSIQ and VQEG video quality databases consisting mainly of compression and transmission based distortions. Another VQA metric that also employs two deep neural networks, 3D CNN and LSTM, is developed in \cite{you2019deep}. 
In addition, the use of Residual Networks (ResNet) for VQA has also been recently investigated. In \cite{li2019quality}, the authors have used pretrained ResNet-50 with global pooling as a spatial feature extractor and called it as content-aware feature extraction. For temporal modeling, they have used Gated Recurrent Unit (GRU) network for learning long-term dependencies and for feature aggregation, followed by subjectively-inspired temporal pooling. They have designed their network specifically for in-the-wild videos and hence did not consider the aspect of distortion classification in their work. 
\vspace{-0.2cm}
\subsection{VQA of medical data}
Quality evaluation of medical data is a much more challenging task as compared to that of natural image and video data. This is due to the differences in imaging modalities used, the type of content and the distortions and the sensitivity in terms of the outcomes for the patient. All the works described above have been devoted only to natural video datasets, and as mentioned earlier, very few works have been developed for VQA of medical data. Indeed, most of them resort to classical IQA and VQA metrics already designed for natural data. Moreover, all of these works focus only either on the quality of video in the context of wireless network transmission and compression  \cite{panayides2011atherosclerotic,martini20133,razaak2014study,usman2019suitability} or in case of IQA only on some specific modalities \cite{amirrashedi2021leveraging, fantini2021automatic, das2021diagnostic, zago2018retinal, wu2017fuiqa}. 
For example, for quality assessment of compressed laparoscopic videos, Kumcu \textit{et al.} \cite{kumcu2014visual} have used average PSNR, reduced-reference Video Quality Metric (VQM) \cite{pinson2004new} and average of frame-based FR high dynamic range visual difference predictor (HDR-VDP-2) metric \cite{mantiuk2011hdr}. On the other hand, Munzer \textit{et al.} \cite{munzer2014investigation} only provide subjective evaluations for compressed laparoscopic videos. In our previous work \cite{khan2020towards}, we evaluated performance of some of the IQA and VQA methods on laparoscopic videos affected by other types of distortion. These metrics include PSNR, SSIM \cite{wang2004image}, VIF \cite{sheikh2004image}, VIIDEO \cite{mittal2015completely}, BRISQUE \cite{mittal2012no} and NIQE \cite{mittal2012making}. Our results illustrated that none of these metrics showed a good correlation with the subjective scores when all the distortions were considered.
\vspace{-0.2cm}
\subsection{Contributions}
Based on the existing literature, it is clear that there is still a big gap when it comes to video quality assessment of laparoscopic videos, especially for distortions other than compression artifacts. Furthermore, the conventional VQA metrics for natural videos have also shown poor correlations with subjective scores for laparoscopic videos~\cite{khan2020towards}. In this paper, we try to fill the gap in laparoscopic video quality monitoring application by proposing an effective method for quality assessment. The main contributions of this paper are listed below: 
\begin{itemize}
	\item First, a new neural network-based approach is developed for NR quality assessment of distorted laparoscopic videos affected by different typical distortions with various severity levels. 
	\item Second, the proposed approach is suitable for a dual task of distortion classification and quality prediction based on a combination of ResNet and Fully Connected Neural Network (FCNN). It is important to emphasize here that designing such architecture for the dual-task of frame-level distortion classification and video quality prediction is extremely important for quality monitoring application. To the best of our knowledge, none of the existing deep learning based VQA methods handle these tasks simultaneously in the medical context.
	\item Third, the effective temporal aggregation based on FCNN proposed in this work is a novel approach that has not been considered in any other works on video quality assessment to the best of our knowledge. Most existing works rely either on inefficient conventional temporal pooling methods or on highly computational and difficult to train recurrent neural networks.
	\item Fourth, the ResNet architecture is firstly trained using data with distortion ranking as labels instead of subjective quality scores \cite{itu2008p910} (that are hard to get and need time). This allows us to overcome the problem of lack of large amounts of labeled video data. Note that this idea of using rankings is inspired by a previous work developed for IQA \cite{liu2017rankiqa}, where the authors employ a Siamese network for ranking distortion severity level before fine-tuning its single trained branch to obtain quality scores. However, unlike \cite{liu2017rankiqa}, where only distortion rank is considered, we perform here the training on data with combined labels corresponding to distortion type and rank.
	\item Fifth, the proposed new architecture performs jointly distortions classification/ranking and severity levels identification.
	\item Finally, the previous pre-trained ResNet model is fine-tuned for both the frame-level distortion classification task as well as for the video quality prediction task while adding an FCNN to produce a final video quality score. For instance, the whole architecture is trained using an end-to-end learning approach. Moreover, in order to compute the video quality score from the frame-level ones, different neural network-based temporal pooling stages have also been investigated.
\end{itemize}  
It should be noted that a preliminary version of this work has been presented in \cite{khan2020residual}. The previous work has been devoted to distortion classification and ranking and has been developed for laparoscopic images. In this paper, we firstly extend this previous work to predict frame-level quality score. Moreover, we added here a second neural network (FCNN) to make the previous method suitable for video quality prediction. In this respect, two learning approaches are investigated to learn the whole architecture. Finally, extensive experiments are conducted here to show the effectiveness of the proposed method with respect to recent conventional and deep learning based video quality assessment methods. 

The remainder of this paper is organized as follows. In Section \ref{sec:icip}, we describe the proposed architecture for frame-level distortion classification and ranking. After that, the extension of the previous architecture to laparoscopic video quality prediction is presented in Section \ref{sec:proposed}. Finally, experimental results are shown and discussed in Section \ref{sec:experiments}, and some conclusions as well as perspectives are drawn in Section \ref{sec:conc}. 

\section{Proposed ResNet-based distortion classification and ranking}
\label{sec:icip}
As mentioned in Section \ref{sec:intro}, in the standard two-stage framework based image quality assessment, the feature extraction step plays a crucial role for distortion classification as well as for quality score prediction. In this section, we will focus on the first stage where a ResNet-based solution is proposed to perform jointly distortion classification and ranking. Before describing our approach, let us first describe our developed Laparoscopic Video Quality (LVQ) database recently introduced in \cite{khan2020towards}.  
\subsection{Overview of LVQ Database}
\label{dataset_descrip}
\begin{figure}[!h]
	\centering
	\subfloat[MB1 \label{fig:mb1}]{\includegraphics[width=0.1\textwidth]{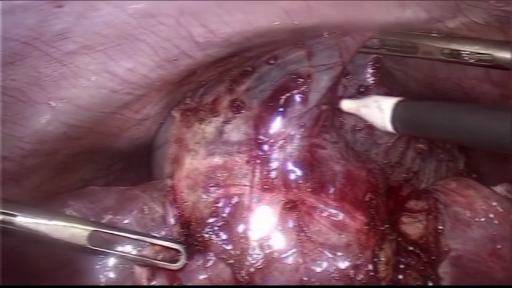}}\hspace{0.01\textwidth}
	\subfloat[MB2 \label{fig:mb2}]{\includegraphics[width=0.1\textwidth]{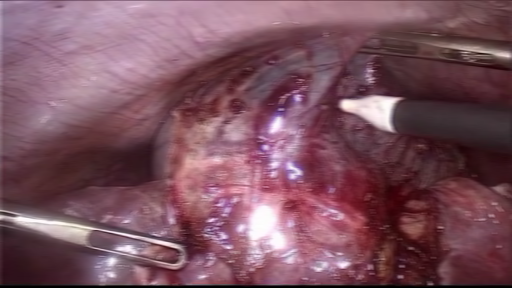}}\hspace{0.01\textwidth}
	\subfloat[MB3 \label{fig:mb3}] {\includegraphics[width=0.1\textwidth]{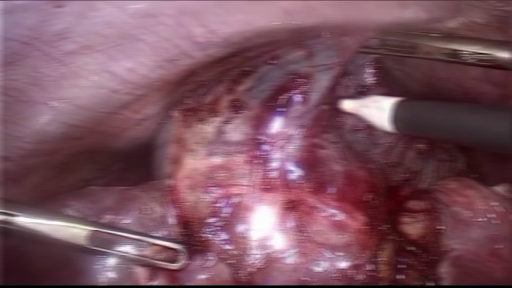}}\hspace{0.01\textwidth}
	\subfloat[MB4 \label{fig:mb4}] {\includegraphics[width=0.1\textwidth]{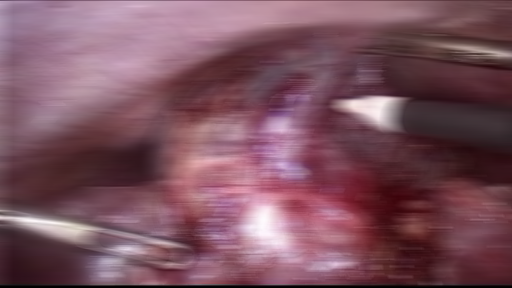}}\hspace{0.01\textwidth}
	\subfloat[DF1 \label{fig:df1}]{\includegraphics[width=0.1\textwidth]{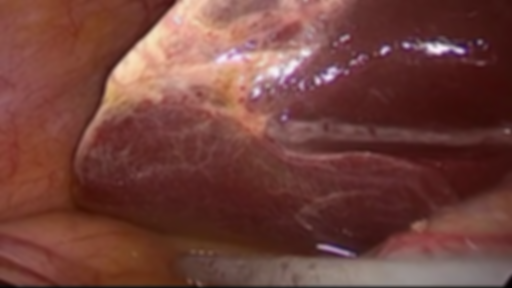}}\hspace{0.01\textwidth}
	\subfloat[DF2 \label{fig:df2}]{\includegraphics[width=0.1\textwidth]{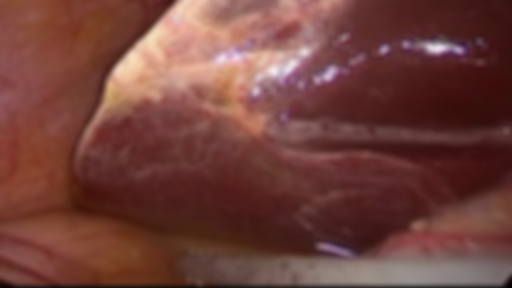}}\hspace{0.01\textwidth}
	\subfloat[DF3 \label{fig:df3}]{\includegraphics[width=0.1\textwidth]{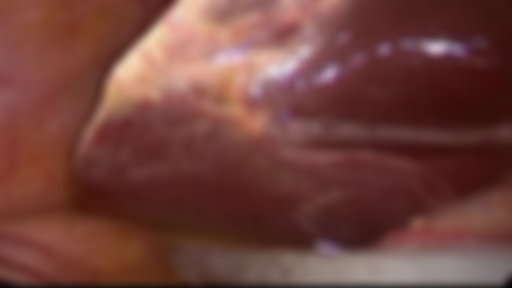}}\hspace{0.01\textwidth}
	\subfloat[DF4 \label{fig:df4}]{\includegraphics[width=0.1\textwidth]{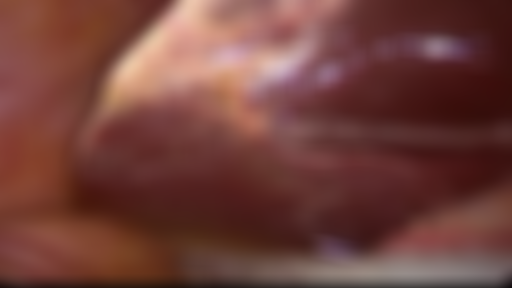}}\hspace{0.01\textwidth}
	\subfloat[UI1\label{fig:ui1}] {\includegraphics[width=0.1\textwidth]{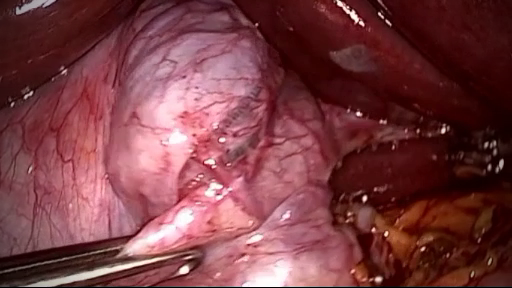}}\hspace{0.01\textwidth}
	\subfloat[UI2\label{fig:ui2}] {\includegraphics[width=0.1\textwidth]{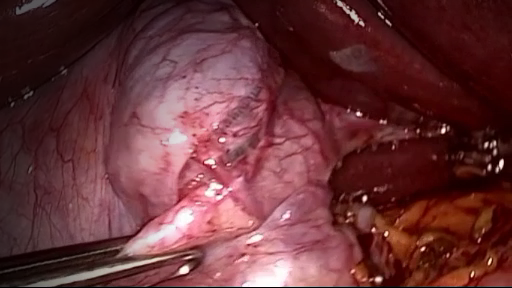}}\hspace{0.01\textwidth}
	\subfloat[UI3\label{fig:ui3}]{\includegraphics[width=0.1\textwidth]{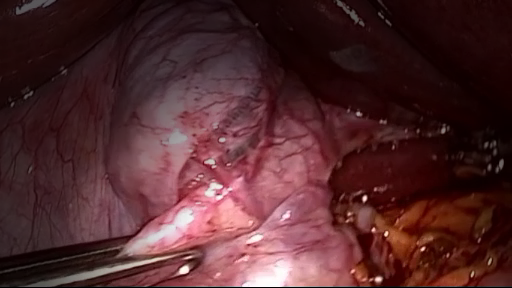}}\hspace{0.01\textwidth}
	\subfloat[UI4\label{fig:ui4}]{\includegraphics[width=0.1\textwidth]{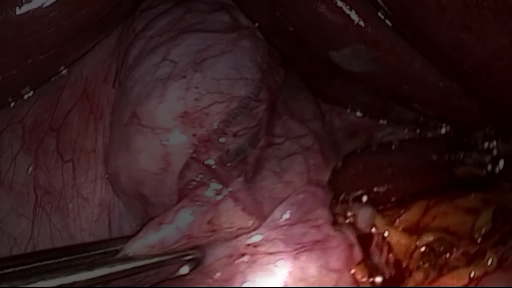}}\hspace{0.01\textwidth}
	\subfloat[SM1 \label{fig:sm1}]{\includegraphics[width=0.1\textwidth]{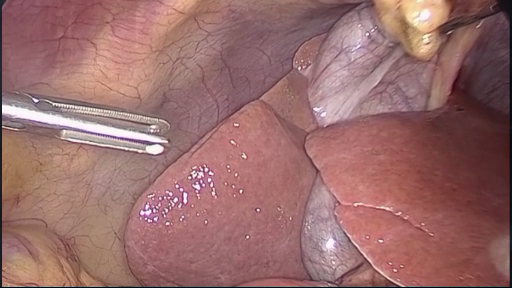}}\hspace{0.01\textwidth}
	\subfloat[SM2 \label{fig:sm2}]{\includegraphics[width=0.1\textwidth]{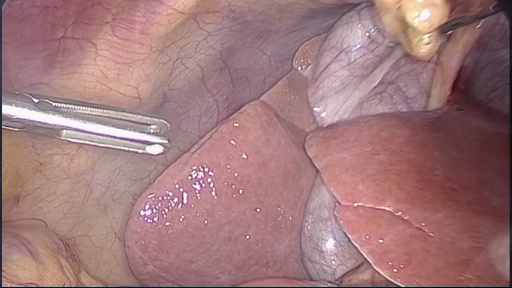}}\hspace{0.01\textwidth}
	\subfloat[SM3\label{fig:sm3}] {\includegraphics[width=0.1\textwidth]{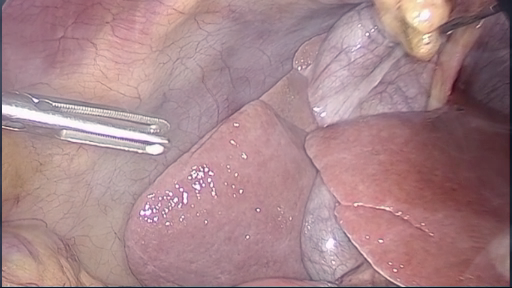}}\hspace{0.01\textwidth}
	\subfloat[SM4\label{fig:sm4}] {\includegraphics[width=0.1\textwidth]{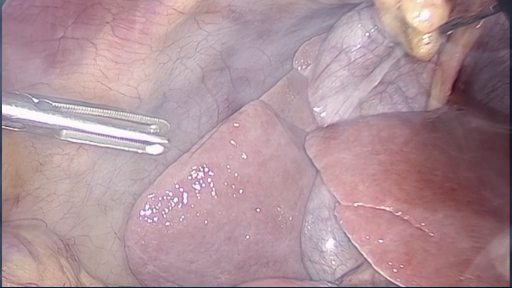}}\hspace{0.01\textwidth}
	\subfloat[WN1 \label{fig:noise1}]{\includegraphics[width=0.1\textwidth]{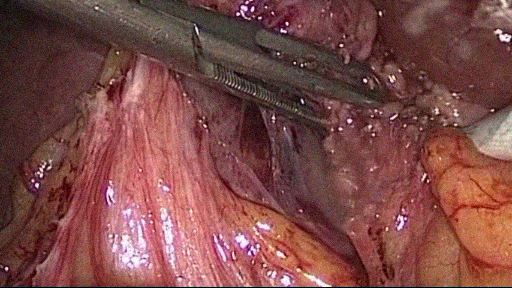}}\hspace{0.01\textwidth}
	\subfloat[WN2 \label{fig:noise2}]{\includegraphics[width=0.1\textwidth]{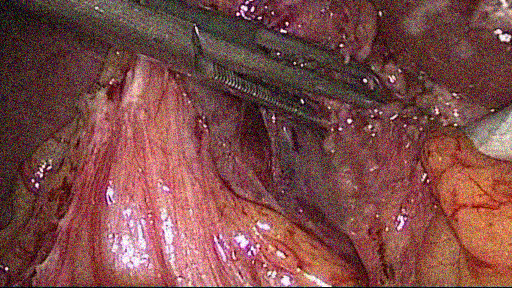}}\hspace{0.01\textwidth}
	\subfloat[WN3 \label{fig:noise3}]{\includegraphics[width=0.1\textwidth]{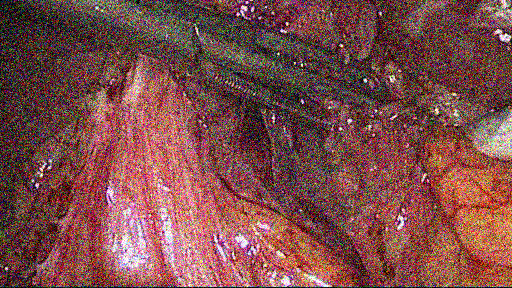}}\hspace{0.01\textwidth}
	\subfloat[WN4 \label{fig:noise4}]{\includegraphics[width=0.1\textwidth]{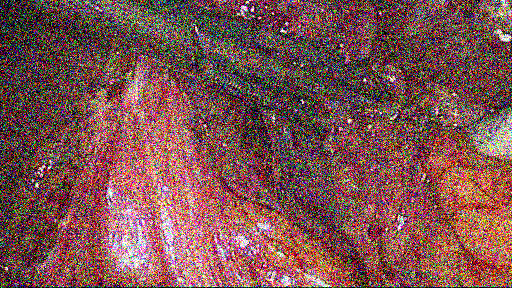}}
	\caption{\label{fig:diff_dist} Illustration of the different distortions at each severity level for representative frames taken from the LVQ dataset.}
	\vspace{-0.5cm} 
\end{figure}

LVQ database is composed of 10 reference videos that have been distorted by five typical distortions at four severity levels, yielding a set of 200 distorted videos. For the reference videos, ten clips of 10 seconds were extracted from different videos of the Cholec80 dataset \cite{twinanda2017endonet} which contains a total of 80 videos of cholecystectomy surgeries. To create the LVQ dataset, the reference videos were selected with an attempt to include maximum possible variations of the scene content and temporal information.
The resolution of the videos in the database is 512 $\times$ 288 with a 16:9 aspect ratio and the frame-rate is 25 fps. \\
The five distortions in the LVQ database, which are among the most common ones affecting laparoscopic video, are the smoke (SM), additive white Gaussian noise (WN), uneven illumination (UI), blur due to defocus (DB) and blur due to motion(MB). For each distortion, four different severity levels have been considered. These levels are labeled, from the least to the most distorted one, as hardly visible (HV), just noticeable (JN), very annoying (VA) and extremely annoying (EA). Numerically, they may be represented by numbers from 1 to 4 with 1 being the least severe and 4 being the most severe. Fig.~\ref{fig:diff_dist} illustrates these different distortions on some representative frames. In order to simulate these distortions and generate the distorted videos, appropriate signal acquisition models were applied to each frame of the video. \\ 
Once all the distorted videos were synthesized, their quality was assessed subjectively by 29 observers. For the subjective testing, the pairwise-comparison protocol
\cite{itu2008p910,qureshi2017towards} was used. More details abo the different distortion models and subjective testing can be found in \cite{khan2020towards}. \\
It is important to note here that, among the medical VQA databases used in the literature, the developed LVQ is the only database to the best of our knowledge, which is made publicly available\footnote{https://github.com/zakopz/LVQ\_Database}. It is also unique in several aspects, such as the diversity of distortions and at different levels, and especially the relatively high participation of expert and non-expert observers. 

\subsection{Problem formulation}

Once the distorted LVQ database is built, and in order to deal with the lack of labeled data, we propose now to aggregate the tasks of distortion classification and quality ranking as a single multi-label classification problem. 
More precisely, let us denote by $X_{d_i,l_j}^{(m)}$ the $m$-th video sequence in the dataset with distortion type $d_i$ and severity level $l_j$, where
\begin{align}
& d_i \in \{DB,MB,WN,SM,UI\} \\
& l_j \in \{HV,JN,VA,EA\}
\end{align}
Such multi-label classification problem can be transformed into a single-label multiclass classification task using one of the Problem Transformation methods \cite{tsoumakas2007multi}. Amongst these methods, we use the Label Powerset approach where each combination of label $(d_i,l_j)$ can be considered as a separate class. This will result in the following set of 20 classes $C$:
\begin{align}
C=\{(d_i,l_j)_{\substack{d_i \in \{DB,MB,WN,SM,UI\} \\l_j \in \{HV,JN,VA,EA\}}}\},
\end{align}
Therefore, the joint distortion classification and ranking tasks can be seen as a single-label multiclass classification problem which will now be solved using a deep learning approach. 
\vspace{-0.3cm}
\subsection{ResNet-based solution} 
While deep convolutional neural networks have been widely investigated in the literature, it has been shown that they may lead to vanishing gradient problem and drop in accuracy performance due to the use of many stacking layers. To solve this problem, Residual Network (ResNet) has been developed by introducing skip connections \cite{he2016deep}. For instance, the latter has been retained in different image processing tasks such as recognition \cite{he2016deep}, classification \cite{TANSY2020}, etc. For this reason, we propose in this paper to resort to the ResNet architecture to solve our distortion classification and ranking problem. 

\begin{figure*}[!h]
    \vspace{-0.8cm}
	\centering
	\includegraphics[width = 0.8\textwidth]{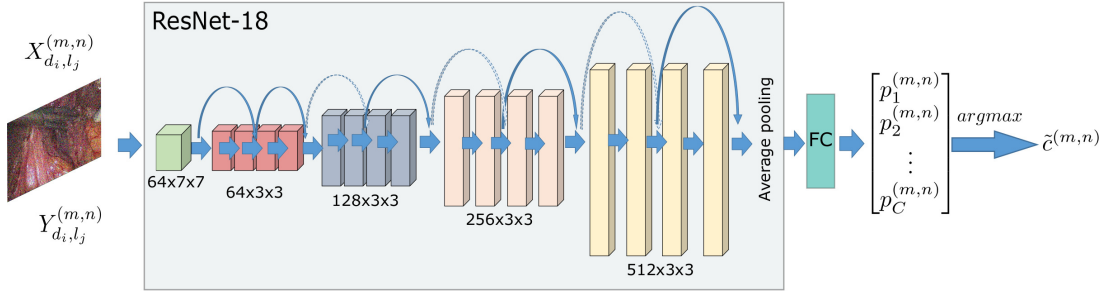}
	\caption{Proposed Frame-level Distortion Classification Residual Network (FDC-ResNet).}
	\label{fig:Resnet_image_class}
	\vspace{-0.5cm}
\end{figure*}

In this respect, and in order to deal with our video database, we first propose to apply separately the ResNet architecture to all the frames of the different video sequences \cite{khan2020residual}. It is important to note here that using such a frame-based approach allows us to overcome the problem of limited labeled training data for videos. The proposed solution, referred to as Frame-level Distortion Classification ResNet (FDC-ResNet), is shown in Fig. \ref{fig:Resnet_image_class}. 

More precisely, as employed in conventional image classification problem, any given variant of ResNet architectures (i.e. ResNet-18, ResNet-34, ResNet-50, etc) is first applied to the different frames of the distorted video, which will be denoted by $X^{(m,n)}_{d_i,l_j}$ where $m \in \{1,\ldots,M\}$ represents the video sequence index in the dataset and $n \in \{1,\ldots,N\}$ refers to the frame index in that video sequence. Let us also assume that $M \times N$ training distorted frames $(X^{(1,1)}_{d_i,l_j},\ldots,X^{(1,N)}_{d_i,l_j},\ldots,X^{(M,1)}_{d_i,l_j},\ldots,X^{(M,N)}_{d_i,l_j})$ with corresponding labels $(Y^{(1,1)}_{d_i,l_j},\ldots,Y^{(1,N)}_{d_i,l_j},\ldots,Y^{(M,1)}_{d_i,l_j},\ldots,Y^{(M,N)}_{d_i,l_j})$ are available. Thus, by performing the convolution and pooling operations on the training samples in a mini-batch, a feature vector is then generated for each sample. Then, a fully connected layer, with output size equal to the number of classes $C$, is used with a softmax function to generate the following vector of probability scores:
\begin{align}
\mathbf{p}^{(m,n)}=[p_1^{(m,n)},\ldots,p_C^{(m,n)}]^{\top}
\end{align}
where $p_c^{(m,n)}$ is the estimated probability that $X^{(m,n)}_{d_i,l_j}$ belongs to the $c$-th class.  

This model is trained by minimizing the cross-entropy function given by
\begin{align}
\mathcal{L}_\mathcal{C}=-\frac{1}{N_\mathcal{B}}\sum_{m,n}\sum_{c=1}^{C} \mathbbm{1}[Y_{d_i,l_j}^{(m,n)}=c]\log(p_c^{(m,n)})
\end{align}
where the indicator function $\mathbbm{1}[Y_{d_i,l_j}^{(m,n)}=c]$ is equal to 1 when the label index $Y_{d_i,l_j}^{(m,n)}$ of the frame $X_{d_i,l_j}^{(m,n)}$ is $c$; otherwise it is equal to 0, and $N_\mathcal{B}$ is the mini-batch size. 

Finally, the predicted class $\tilde{c}^{(m,n)}$ of the input frame  $X_{d_i,l_j}^{(m,n)}$ is obtained by selecting the class yielding the maximum probability value:
\begin{align}
\tilde{c}^{(m,n)}=\argmax_{c \in \{1,\ldots,C\}}(p_c^{(m,n)})
\end{align}

Once the FDC-ResNet model is trained, it can be applied on the test video data. Indeed, for each input distorted video $X_{d_i,l_j}^{(m)}$, the trained FDC-ResNet model is applied to all the frames $X_{d_i,l_j}^{(m,n)}$ to generate their respective predicted classes $\tilde{c}^{(m,n)}$, with $n \in \{1,\ldots,N\}$. The trained network is then finally fine-tuned for only distortion classification using $d_i$ as the set of labels. Moreover, the same training can be used for video quality prediction network, proposed in the next section. 

\section{Extension to Laparoscopic Video Quality Prediction}
\label{sec:proposed}

\begin{figure*}[!h]
	\centering
	\includegraphics[width = 0.7\textwidth]{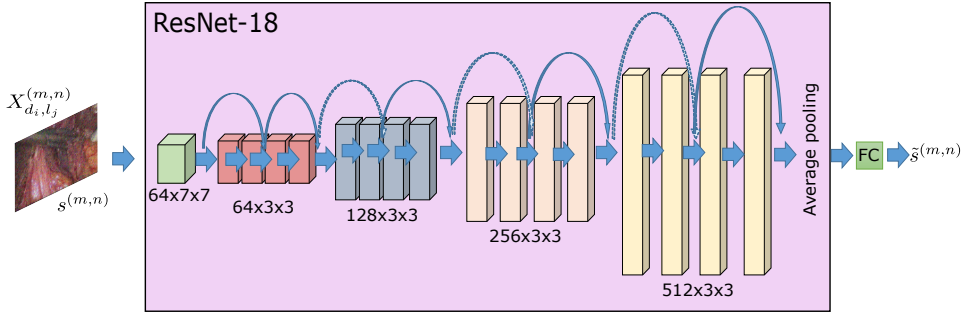}
	\caption{Proposed  Frame-level Quality Prediction Residual Network (FQP-ResNet).}
	\label{fig:Resnet_image_pred}
	\vspace{-0.5cm}
\end{figure*}

\subsection{Motivation}
Once the feature extraction step is done for distortion classification and ranking, a straightforward approach may consist in resorting to a regression module to map the extracted feature vector to a quality score as generally performed in two-stage framework for blind image quality assessment \cite{moorthy2010two}. However, it would be more interesting to adapt the network model and make it more appropriate for quality prediction task, while taking into account the temporal effects. To this end, and similarly to the previous classification step, we first propose to perform a ResNet-based quality prediction stage on the different frames and then merge the resulting frame quality scores, based on a Fully Connected Neural Network (FCNN), to generate the final quality prediction score of a given video. 
\vspace{-0.6cm}
\subsection{Modified ResNet-based solution}
By following the same strategy used in the previous task, and in order to deal with the limited number of labeled training data for videos, we first propose to apply separately our ResNet-based solution on the different frames of the dataset. More precisely, the modified ResNet-based solution, referred to as Frame-level Quality Prediction Residual Network (FQP-ResNet), is shown in 
Fig.~\ref{fig:Resnet_image_pred}.

Thus, compared to the previous FDC-ResNet, and after generating the feature vector using the convolution and pooling layers, a fully connected layer of size one neuron in used in the FQP-ResNet to produce the predicted score $\tilde{s}^{(m,n)}$ associated to the $n$-th frame of the $m$-th distorted video. For training this modified ResNet-based prediction model, we propose to start from the pre-trained ResNet model used for classification (as an initialization step) and then fine-tune its different weights to adapt it to the quality prediction task. As a good quality assessment metric should be well correlated with human opinion, the loss function used to optimize the FQP-ResNet weight parameters will be defined based on the Pearson correlation coefficient often employed as a standard criterion to compare image/video quality assessment methods. To this end, we propose to minimize the following loss function:  
\begin{align}
\widetilde{\mathcal{L}}_\mathcal{P}=1-\frac{\sum_{m,n}\Big(s^{(m,n)}-\mu_{s}\Big) \Big(\tilde{s}^{(m,n)}-\mu_{\tilde{s}}\Big)}{\sqrt{\sum_{m,n}\Big(s^{(m,n)}-\mu_{s}\Big)^2\sum_{m,n}\Big(\tilde{s}^{(m,n)}-\mu_{\tilde{s}}\Big)^2}}
\end{align}
where $s^{(m,n)}$ is the target subjective score associated to the $n$-th frame of the $m$-th distorted video and, $\mu_{s}$ and $\mu_{\tilde{s}}$ are the means of the target subjective scores and the predicted ones, respectively. 

It should be noted here that the subjective score, obtained during the subjective test, are provided for the overall video sequence. Hence, we will assume here that the subjective scores of the different frames are equal to that of the corresponding video sequence. Despite its simplicity, this assumption is in some way reasonable for the following two reasons. First, in the context of laparoscopic surgery, typical video sequences do not show significant variations between the frame contents. Moreover, the temporal effect aspect will be taken into account later when merging the frame quality scores to produce a final score for the overall video.

\begin{figure*}[!h]
	\centering
	\includegraphics[width=0.85\textwidth]{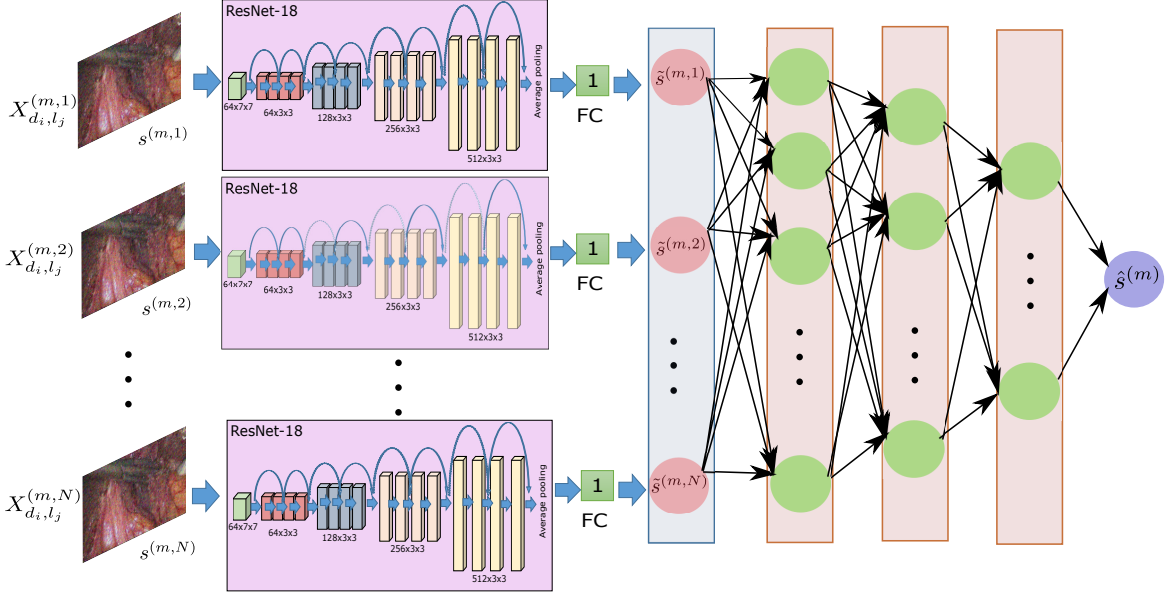}
	\caption{Proposed Video Quality Prediction Network (VQP-Net).}
	\label{fig:Resnet_video_pred}
\end{figure*}

Once the FQP-ResNet model is described, we will focus now on its application for video quality prediction. The complete architecture, referred to as Video Quality Prediction Network (VQP-Net), is illustrated in Fig.~\ref{fig:Resnet_video_pred}.

As it can be seen from Fig.~\ref{fig:Resnet_video_pred}, for any input distorted video sequence $X^{(m)}_{d_i,l_j}$, the previous FQP-ResNet is first separately applied to the different frames $X^{(m,n)}_{d_i,l_j}$ resulting in the following subjective quality score vector $\tilde{\mathbf{s}}^{(m)}$:
\begin{align}
\tilde{\mathbf{s}}^{(m)}=[\tilde{s}^{(m,1)},\ldots,\tilde{s}^{(m,N)}]^{\top}
\end{align}
Then, in order to take into account the temporal effect, an additional FCNN is incorporated to combine the different frame quality scores and produce a final video quality score. More precisely, the subjective quality score vector $\tilde{\mathbf{s}}^{(m)}$ is associated to the input layer of this FCNN. Then, $H$ hidden layers are used where their neuron values are computed from the previous ones based on a linear combination (with bias) followed by a nonlinear activation function. Finally, an output layer with a single neuron is employed yielding the computation of the predicted quality score $\hat{s}^{(m)}$ associated to the input distorted video sequence $X^{(m)}_{d_i,l_j}$. \\
For the training of this proposed VQP-Net model, two approaches will now be addressed. 

\subsubsection{Transfer learning approach} 
One straightforward approach would consist in following a process similar to the transfer learning which is based on the use of a pre-trained model. Thus, the previous pre-trained FQP-ResNet model will be firstly used by the first block of the overall architecture (see Fig.~\ref{fig:Resnet_video_pred}) to generate the frame quality scores. Then, our training strategy will focus only the learning of the FCNN weight parameters. This will be achieved by maximizing the correlation between the target subjective score $s^{(m)}$ and the predicted one $\hat{s}^{(m)}$. Therefore, the FCNN weight parameters are updated by minimizing the following Pearson correlation coefficient based loss function: 
\begin{align}
\mathcal{L}_\mathcal{P}=1-\frac{\sum_{m}\Big(s^{(m)}-\mu_{s}\Big) \Big(\hat{s}^{(m)}-\mu_{\hat{s}}\Big)}{\sqrt{\sum_{m}\Big(s^{(m)}-\mu_{s}\Big)^2\sum_{m}\Big(\hat{s}^{(m)}-\mu_{\hat{s}}\Big)^2}}
\label{eq:VQP}
\end{align}

Once the FCNN model is trained, the obtained learned parameters as well as those of the pre-trained ResNet one are employed by the complete VQP-Net architecture for computing the prediction quality score of each input test video sequence. 

\subsubsection{End-to-end learning approach}
While the previous approach presents the advantage of reducing the complexity of the training phase, a more interesting approach should consist in resorting to an end-to-end learning approach. More precisely, instead of using the pre-trained FQP-ResNet model (while keeping it fixed) in the first block of the VQP-Net architecture, we propose here to update the weight parameters of the FQP-ResNet model during the training phase. Therefore, the parameters of both the FQP-ResNet model (i.e first block) and the FCNN one (i.e second block) of our VQP-Net architecture are simultaneously optimized by using the same loss function defined in \eqref{eq:VQP}. The final learned parameters of the whole VQP-Net model are then used to predict the quality scores of the test distorted videos. 

\begin{figure}[!h]
	\centering
	\includegraphics[width=0.35\textwidth]{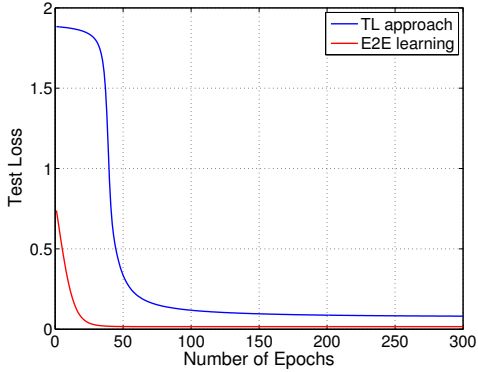}
	\caption{Loss function evolution with the number of epochs for the transfer learning (TL) and end-to-end (E2E) learning approaches.}
	\label{fig:train_comparison}
	\vspace{-0.5cm}
\end{figure}

To show the interest of this end-to-end learning approach compared to the transfer learning one, Fig. \ref{fig:train_comparison} illustrates the loss functions (given by \eqref{eq:VQP}) evaluated on the validation dataset with respect to the number of epochs. Thus, it can be observed the end-to-end training approach outperforms the transfer learning one in terms of loss function values. Moreover, the convergence of the  end-to-end training approach is much faster than that of the transfer learning technique.

\section{Experimental Results}
\label{sec:experiments}
In this section, intensive experiments have been conducted to evaluate the performance of the proposed video distortion classification and quality prediction methods. In the following, we will define the experimental settings used in our simulations, present the comparison methods and finally discuss the obtained results.

\vspace{-0.3cm}
\subsection{Experimental settings}

The proposed ResNet based methods were tested on the LVQ dataset described in Section \ref{dataset_descrip}. For the task related to the distortion classification and ranking (FDC-ResNet), the original training dataset has been extended by creating more distorted videos using the Cholec80 dataset \cite{twinanda2017endonet} and following the methodology explained in \cite{khan2020towards}. Furthermore, a frame-level data augmentation step is also performed by applying random cropping and random horizontal flipping to the original frames. It is important to note here that we did not carry out any subjective test for these new videos since the subjective scores are not used for training the FDC-ResNet. Indeed, the latter only requires the information about the rank and distortion type. After that, for the quality prediction task (FQP-ResNet), the obtained FDC-ResNet model is fine tuned and trained on the original training dataset with available subjective scores. Regarding the VQP-Net task, the FCNN employed in the second block of the overall architecture is implemented using three hidden layers (i.e $H=3$) and the log softmax function as an activation function. Note that 80\% of the original LVQ database is used for training and 20\% for testing. \\
The implementation of the proposed network was done using PyTorch and the network was run on a Windows system with Nvidia Quadro RTX-6000 GPU and 32 GB RAM. For the training of FDC-ResNet, we used the Adam optimizer with a learning rate of 0.01. The learning rate was dynamically reduced if the loss value did not change for two consecutive iterations. For the training of VQP-Net, a lower learning rate of 0.00001 was used with the Adam optimizer.

\subsection{Comparison methods and performance evaluation}
To demonstrate the effectiveness of the proposed video objective quality prediction method, a comparative study with some state-of-the-art VQA metrics is conducted. A first way to evaluate the quality of videos is to consider state-of-the-art IQA metrics designed for images, such as SSIM \cite{wang2004image}, VIF \cite{sheikh2004image}, BRISQUE \cite{mittal2012no} and NIQE \cite{mittal2012making}, apply them frame by frame and use a temporal pooling model to derive VQA measure of the whole video stream. This simple approach was used as the first intuitive solution for estimating the global video quality. For these metrics, we have applied four different temporal pooling techniques for combining the predicted scores from all the frames namely the average pooling, geometric mean pooling, harmonic mean pooling and median pooling. We also considered quality metrics designed specifically for video and which integrate, in one way or another, the temporal aspect according to some pooling models in an explicit or implicit way. These VQA measures can be classified into conventional and deep learning (DL)  based metrics. The conventional metrics used for comparison are V-BLIINDS \cite{saad2014blind}, VIIDEO \cite{mittal2015completely} and TLVQM \cite{korhonen2019two}. For DL based methods, we tested four recent approaches namely Inceptionv3-FT \cite{varga2019no}, VSFA \cite{li2019quality}, CNN-LSTM TLVQM \cite{korhonen2020blind} and CNN-SVR TLVQM \cite{korhonen2020blind}. It should be noted here that these methods have been tested using the source codes provided by the authors except for the Inceptionv3-FT method which is tested using the corrected implementation \cite{gotz2019no}. \\
It is worth mentioning here that all the retained state-of-the-art learning-based methods have also been trained on the new LVQ database for fair comparison.\\
The performance of the different video quality prediction methods is evaluated in terms of various standard criteria measuring the correlation between the predicted scores and the subjective ones. More precisely, we have computed the Pearson Linear Correlation Coefficient (PLCC), Spearman Rank-Order Correlation Coefficient (SROCC) and Kendall Rank-Order Correlation Coefficient (KROCC) after performing five-parameter logistic regression for the predicted scores.
\begin{figure}[!h]
	\centering
	\includegraphics[width=0.45\textwidth]{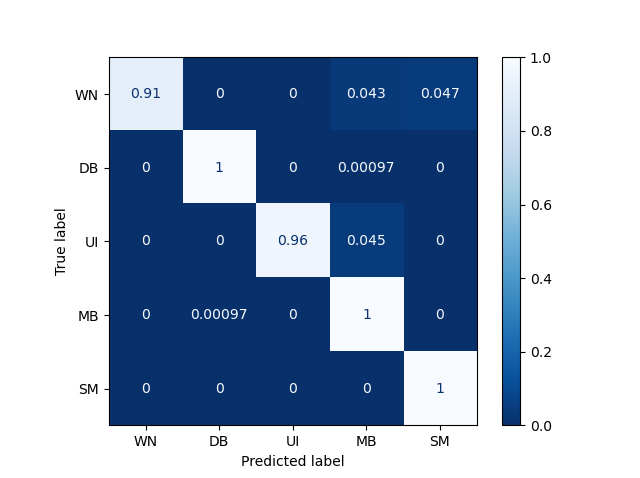}
	\caption{Confusion matrix for the proposed FDC-ResNet.}
	\label{fig:confusion matrix}
	\vspace{-0.5cm}
\end{figure}
\vspace{-0.2cm}
\subsection{Results and discussion}
\textbf{Distortion classification performance}: First, we have evaluated the classification performance of our proposed architecture and have compared it with some existing classification methods. In this regards, we have selected some of the conventional IQA metrics based on the two-stage network which also perform the classification namely BRISQUE, BIQI and BLIINDS-II. Table \ref{tab:class} shows the accuracy results of these methods. Thus, it can be noticed that our proposed method outperforms the conventional ones and yields a classification accuracy of 97.8\%. Furthermore, Fig. \ref{fig:confusion matrix} illustrates well, through the confusion matrix, the superiority of the proposed distortion classification method in terms of prediction error and more particularly in the case of the two most common distortions, i.e. noise and and uneven illumination.

\begin{table}[htbp]
	\caption{Comparison of distortion classification accuracy.}
	\label{tab:class}
	\begin{center}
		\begin{tabular}[width=0.4\textwidth]{|c|c|}
		\hline
		\textbf{Method} & \textbf{Classification Accuracy} \\
		\hline
		\textbf{FDC-ResNet} & 97.8\%\\
		\hline
		\textbf{BRISQUE} & 94.2\%\\
		\hline
		\textbf{BIQI} & 95.0\%\\
		\hline
		\textbf{BLIINDS-II} & 95.6\% \\
		\hline
		\end{tabular}
	\end{center}
	\vspace{-0.3cm}
\end{table}

\textbf{Video quality prediction performance}: Table~\ref{tab:state_of_the_art} shows the results of our proposed methods as well as the state-of-the-art ones for the LVQ database in terms of PLCC, SROCC and KROCC. Note that the category of metric to which belongs each metric (conventional or deep learning (DL) one) as well as its type (full-reference (FR) or no-reference (NR) one) are also provided.
\begin{table*}[htbp]
	\caption{PLCC, SROCC and KROCC for the scores of the different video quality prediction methods.}
	\begin{center}
		\renewcommand\arraystretch{1.2}
		\begin{tabular}[width=0.4\textwidth]{|l|l|l|l|l|l|l|l|l|l|l|l|l|}
			\hline
			\rule[-1ex]{0pt}{2.5ex} \textbf{Metric} & \textbf{Category} & \textbf{Type}  & \textbf{Temporal Model}
			&  \textbf{PLCC} &  \textbf{SROCC} &  \textbf{KROCC} \\
			\hline					
			\rule[-1ex]{0pt}{3.5ex} \multirow{4}*{\textbf{PSNR}}&\multirow{4}*{Conventional}&\multirow{4}*{FR-IQA}&Average Pooling&0.6973&0.6996&0.5030 \\
			\cline{4-7} &  &  & Geometric Mean Pooling & 0.6900 & 0.6949&0.4965 \\
			\cline{4-7} &  &  & Harmonic Mean Pooling & 0.6778 & 0.6829&0.4875 \\
			\cline{4-7} &  &  & Median Pooling & 0.7131 & 0.7097&0.5133 \\	
			\hline					
			\rule[-1ex]{0pt}{3.5ex} \multirow{4}*{\textbf{SSIM \cite{wang2004image}}}& \multirow{4}*{Conventional}& \multirow{4}*{FR-IQA}& Average Pooling&0.6123&0.5914&0.4187 \\	
			\cline{4-7} &  &  & Geometric Mean Pooling & 0.6034 & 0.5812&0.4123 \\
			\cline{4-7} &  &  & Harmonic Mean Pooling & 0.5902 & 0.5563 &0.3980\\
			\cline{4-7} &  &  & Median Pooling & 0.6157 & 0.5945 &0.4183\\
			\hline					
			\rule[-1ex]{0pt}{3.5ex} \multirow{4}*{\textbf{VIF \cite{sheikh2004image}}}&\multirow{4}*{Conventional}&\multirow{4}*{FR-IQA}&Average Pooling&0.6267&0.6228&0.4537 \\	
			\cline{4-7} &  &  & Geometric Mean Pooling & 0.6158 & 0.6307 & 0.4503 \\
			\cline{4-7} &  &  & Harmonic Mean Pooling & 0.5834 & 0.6081 & 0.4397\\
			\cline{4-7} &  &  & Median Pooling & 0.6211 & 0.6192 & 0.4504\\
			\hline
			\rule[-1ex]{0pt}{3.5ex} \multirow{4}*{\textbf{BRISQUE \cite{mittal2012no}}}&\multirow{4}*{Conventional} &\multirow{4}*{NR-IQA}&Average Pooling&  0.4593 &  0.4304&0.3108 \\
			\cline{4-7} &  &  & Geometric Mean Pooling & 0.4609 & 0.4300&0.3114 \\
			\cline{4-7} &  &  & Harmonic Mean Pooling & 0.4722 & 0.4360&0.3199 \\
			\cline{4-7} &  &  & Median Pooling & 0.4639 & 0.4310&0.3119 \\	
			\hline	
			\rule[-1ex]{0pt}{3.5ex} \multirow{4}*{\textbf{NIQE \cite{mittal2012making}}}&\multirow{4}*{Conventional}& \multirow{4}*{NR-IQA}&Average Pooling&0.4242 &   0.3731 &0.3555\\	
			\cline{4-7} &  &  & Geometric Mean Pooling & 0.4535 & 0.4958&0.3594 \\
			\cline{4-7} &  &  & Harmonic Mean Pooling & 0.4402 & 0.4856&0.3641 \\
			\cline{4-7} &  &  & Median Pooling & 0.4583 & 0.4994&0.3633 \\	
			\hline 	
			\hline	
			\rule[-1ex]{0pt}{3.5ex} \textbf{VIIDEO \cite{mittal2015completely}}&Conventional&NR-VQA&Temporal Features&0.3842&0.3416&0.2313
			\\
			\hline
			\rule[-1ex]{0pt}{3.5ex} \textbf{V-BLIINDS \cite{saad2014blind}}&Conventional&NR-VQA&Temporal Features&0.8328&0.8242&0.6317
			\\
			\hline
			\rule[-1ex]{0pt}{3.5ex} \textbf{TLVQM \cite{korhonen2019two}}&Conventional&NR-VQA& Average Pooling&0.7681&0.6892&0.5132
			\\
			\hline
			\hline 
			\rule[-1ex]{0pt}{3.5ex} \textbf{CNN-SVR TLVQM \cite{korhonen2020blind}}&DL&NR-VQA&SVR&0.5826&0.5888&0.4088
			\\
			\hline
			\rule[-1ex]{0pt}{3.5ex} \textbf{CNN-LSTM TLVQM \cite{korhonen2020blind}}&DL&NR-VQA&LSTM&0.8006&0.7829&0.5828
			\\
			\hline
			\rule[-1ex]{0pt}{3.5ex} \textbf{Inceptionv3-FT \cite{varga2019no} \cite{gotz2020comment}}&DL&NR-VQA& Average Pooling&0.8550&0.7978&0.6202
			\\
			\hline 
			\rule[-1ex]{0pt}{3.5ex} \textbf{VSFA \cite{li2019quality}}&DL&NR-VQA& GRU + Subjectively-inspired&0.9647&0.9247&0.7629
			\\	
			\hline
			\rule[-1ex]{0pt}{3.5ex} \textbf{Proposed VQP-Net (TL)}&DL&NR-VQA&Fully-connected Network&0.8992&0.8434&0.6494
			\\	
			\hline
			\rule[-1ex]{0pt}{3.5ex} \textbf{Proposed VQP-Net (E2E)}&DL&NR-VQA&Fully-connected Network&\textbf{0.9899}&\textbf{0.9388}&\textbf{0.7739}
			\\	
			\hline	
		\end{tabular}
		\label{tab:state_of_the_art}
	\end{center}
	\vspace{-0.5cm}
\end{table*}
From the table, it can be firstly observed that the IQA based methods perform poorly  in terms of PLCC, SROCC and KROCC, even when using temporal pooling strategies other than the average pooling one. Among these IQA-based video quality assessment methods, the highest correlation values are obtained using the PSNR metric with median pooling. Secondly, regarding the NR-VQA methods, except for the VIIDEO metric, TLVQM and V-BLIINDS ones outperform all the conventional IQA-based video quality metrics. Indeed, V-BLIINDS leads to the best results compared to the other conventional IQA-based metrics and VQA ones. The obtained PLCC, SROCC and KROCC values are 0.8328, 0.8242 and 0.6317, respectively. Finally, concerning the DL based VQA methods, one can observe that the performance of the CNN-TLVQP method depends on the temporal pooling strategy. Indeed, while the Support Vector Regression (SVR) based temporal pooling stage is less performant compared to some conventional methods, the Long Short-Term Memory (LSTM) based temporal stage improves significantly the results yielding a gain of around 0.2 in all the correlation values. Further improvements are achieved by the Inception-v3 method, and more specifically, the VSFA method. For the proposed VQP-Net method, it can be firstly noticed that its first version based on the transfer learning (TL), outperforms the state-of-the-art deep learning ones except the VSFA method. Most importantly, the end-to-end learned version (E2E) allows us to improve the VSFA method while achieving a gain of around 2.5 \% in terms of PLCC and 1.5~\% in terms of SROCC and KROCC.

\begin{figure}[!http]
	\centering
	\subfloat[PSNR \label{fig:psnr}]{\includegraphics[width=0.15\textwidth]{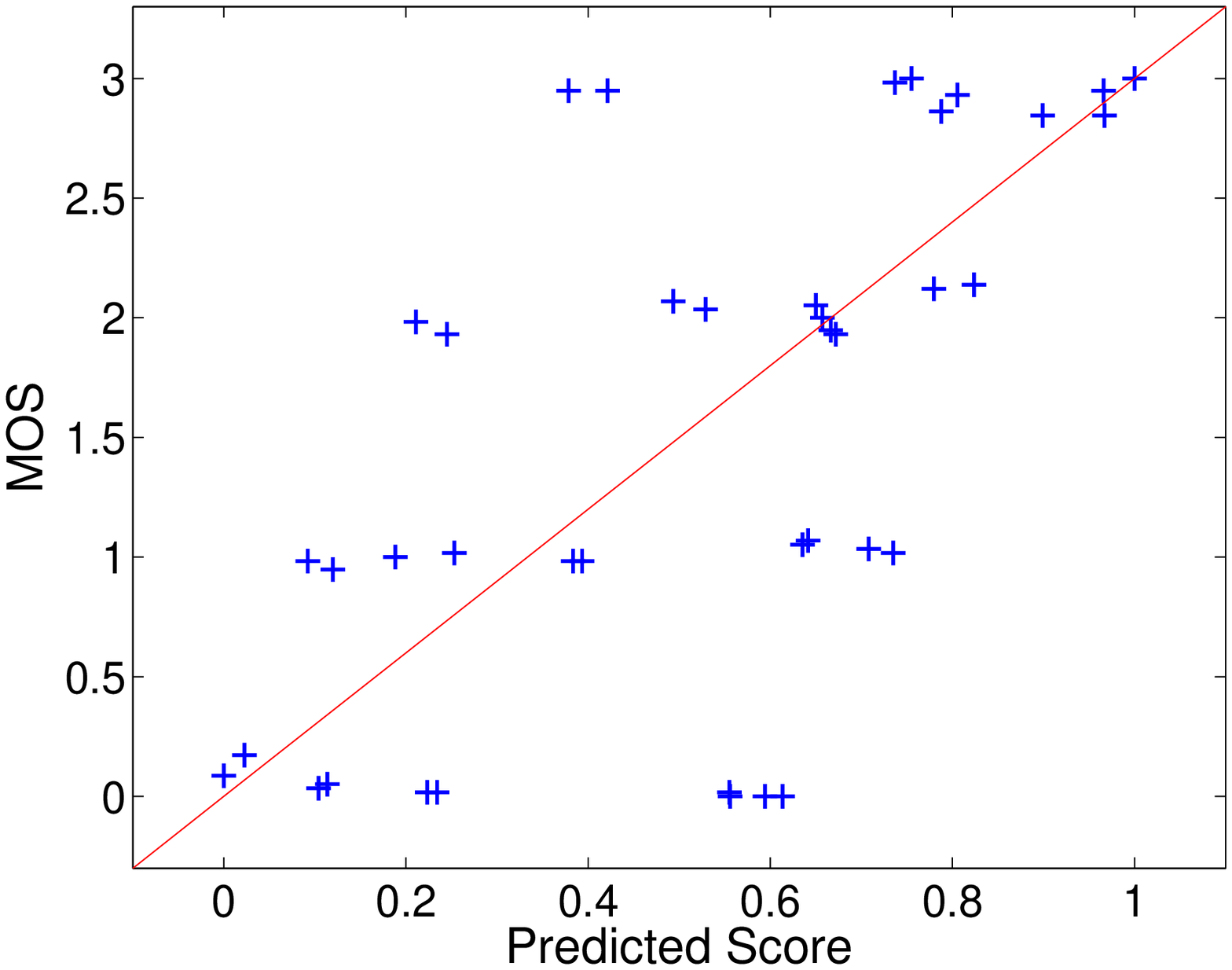}}\hspace{0.01\textwidth}
	\subfloat[SSIM \label{fig:ssim}] {\includegraphics[width=0.15\textwidth]{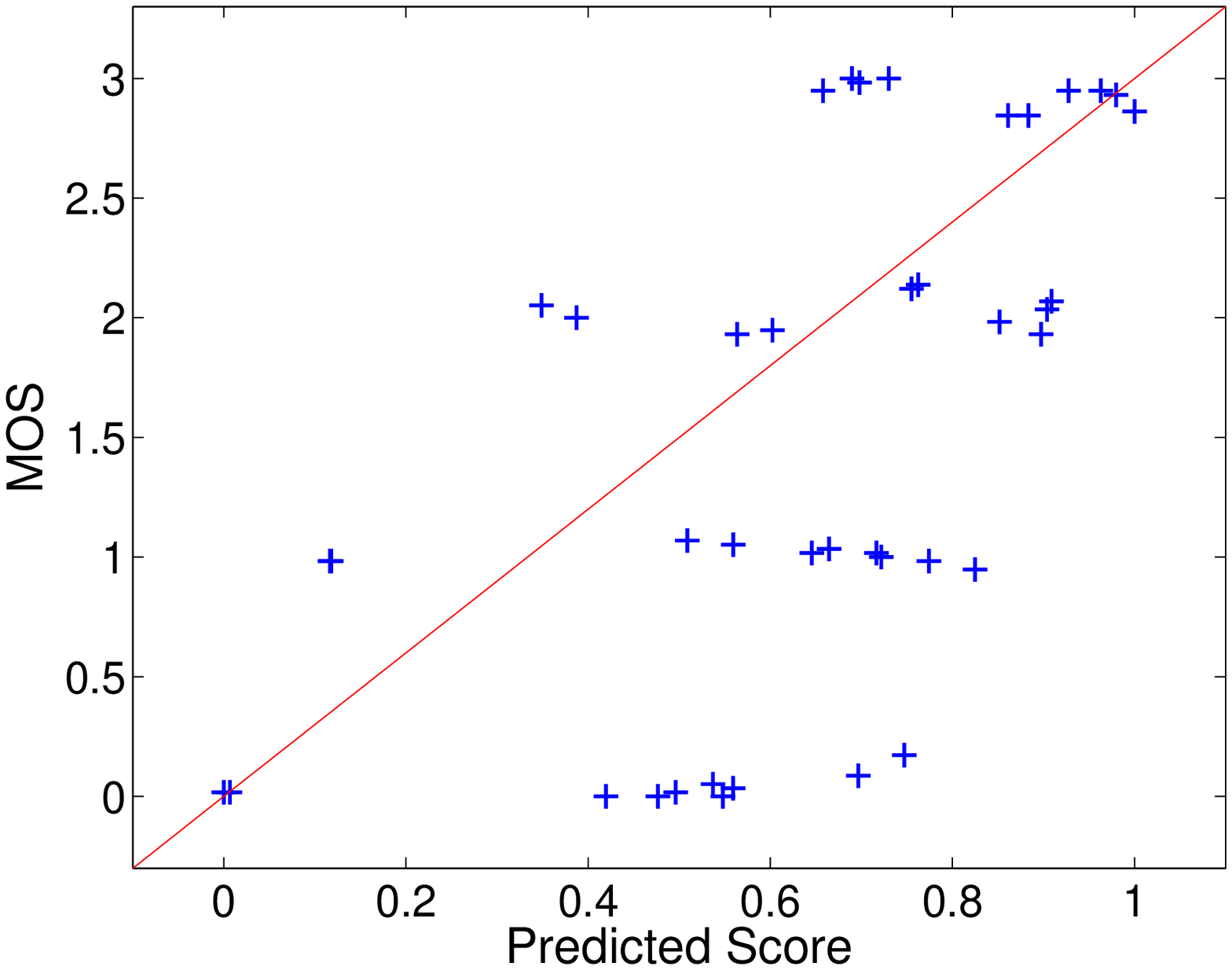}}\hspace{0.01\textwidth}
	\subfloat[VIF \label{fig:vif}]{\includegraphics[width=0.15\textwidth]{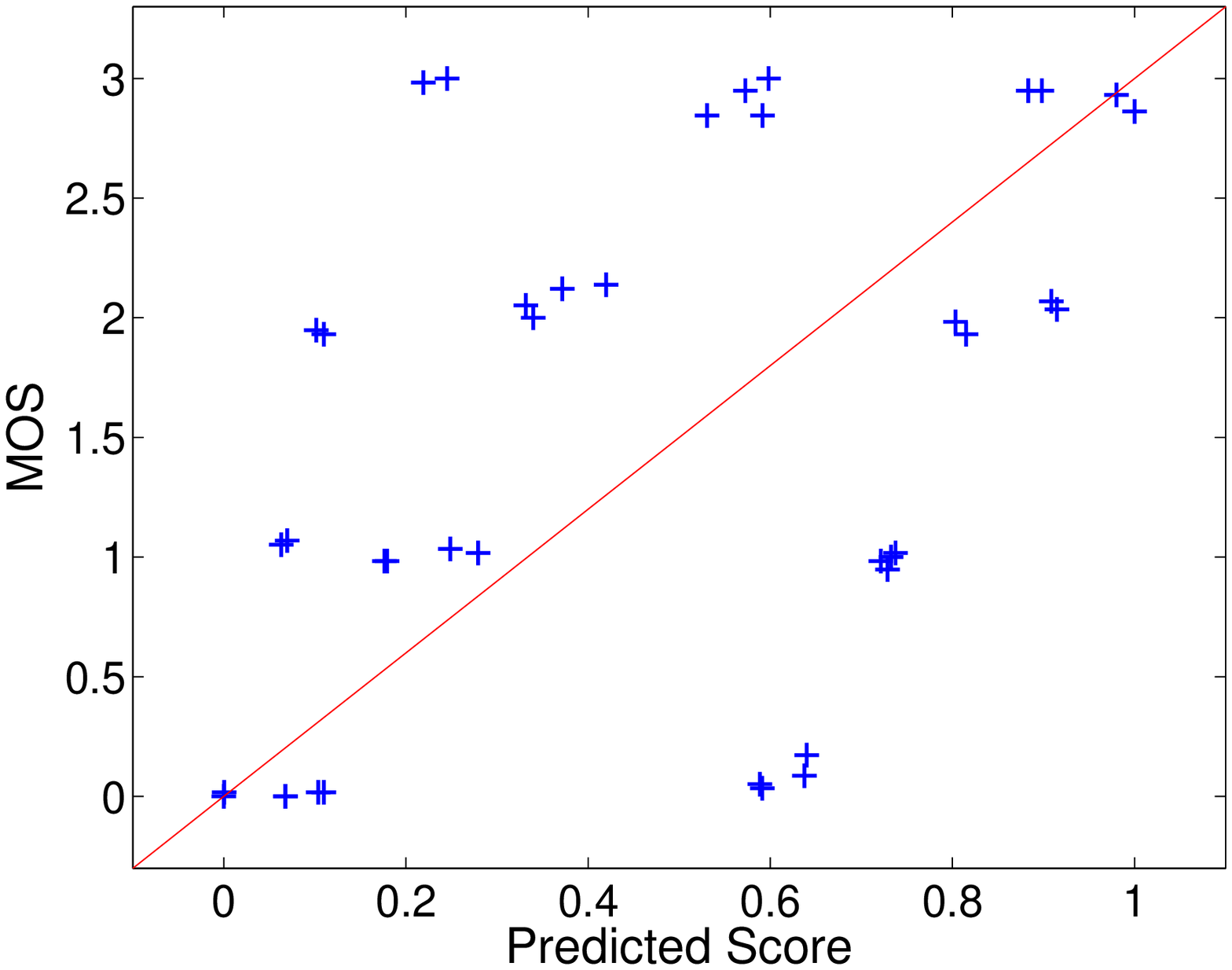}}\hspace{0.01\textwidth}
	\subfloat[BRISQUE \label{fig:brisque}]{\includegraphics[width=0.15\textwidth]{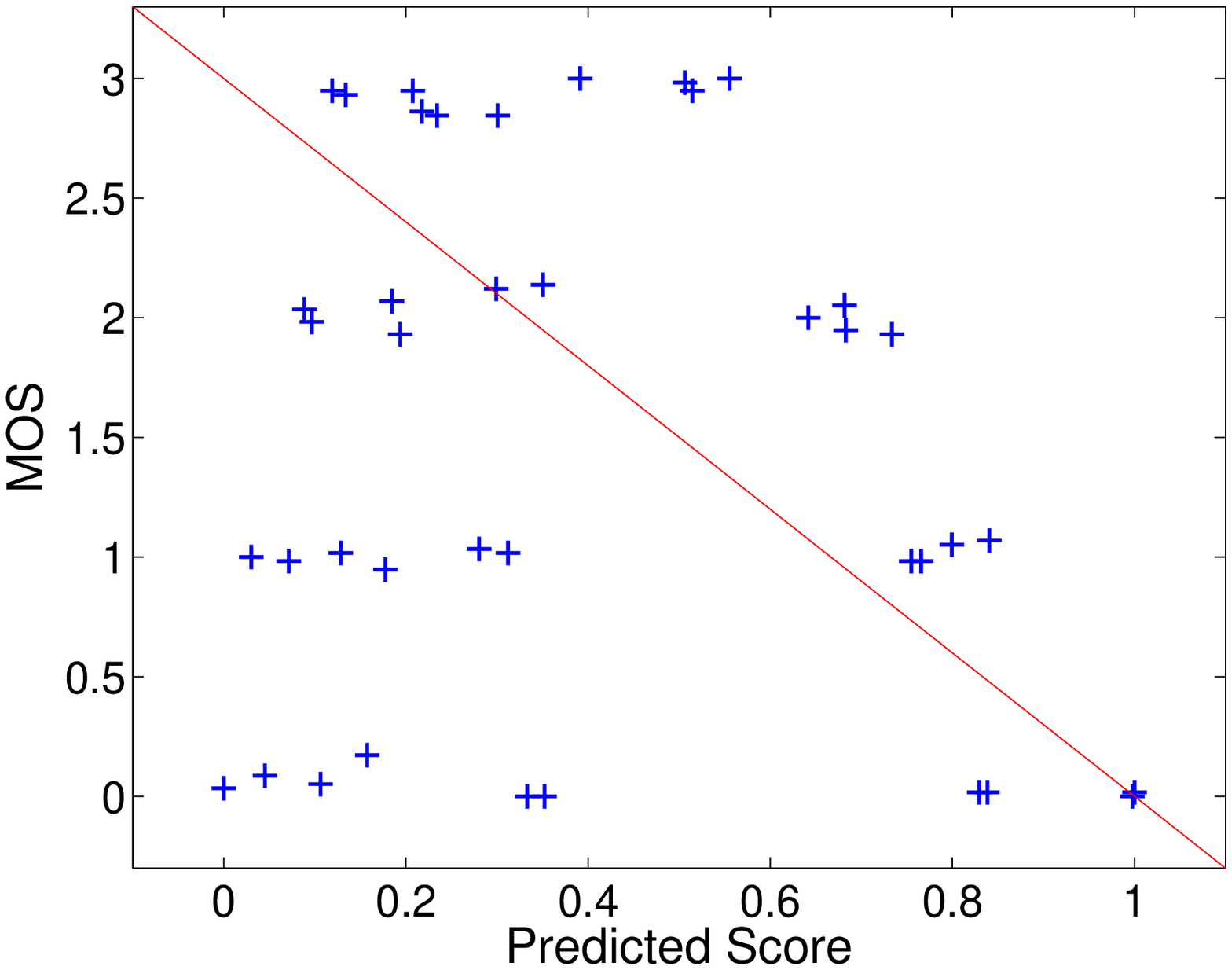}}\hspace{0.01\textwidth}
	\subfloat[NIQE\label{fig:niqe}] {\includegraphics[width=0.15\textwidth]{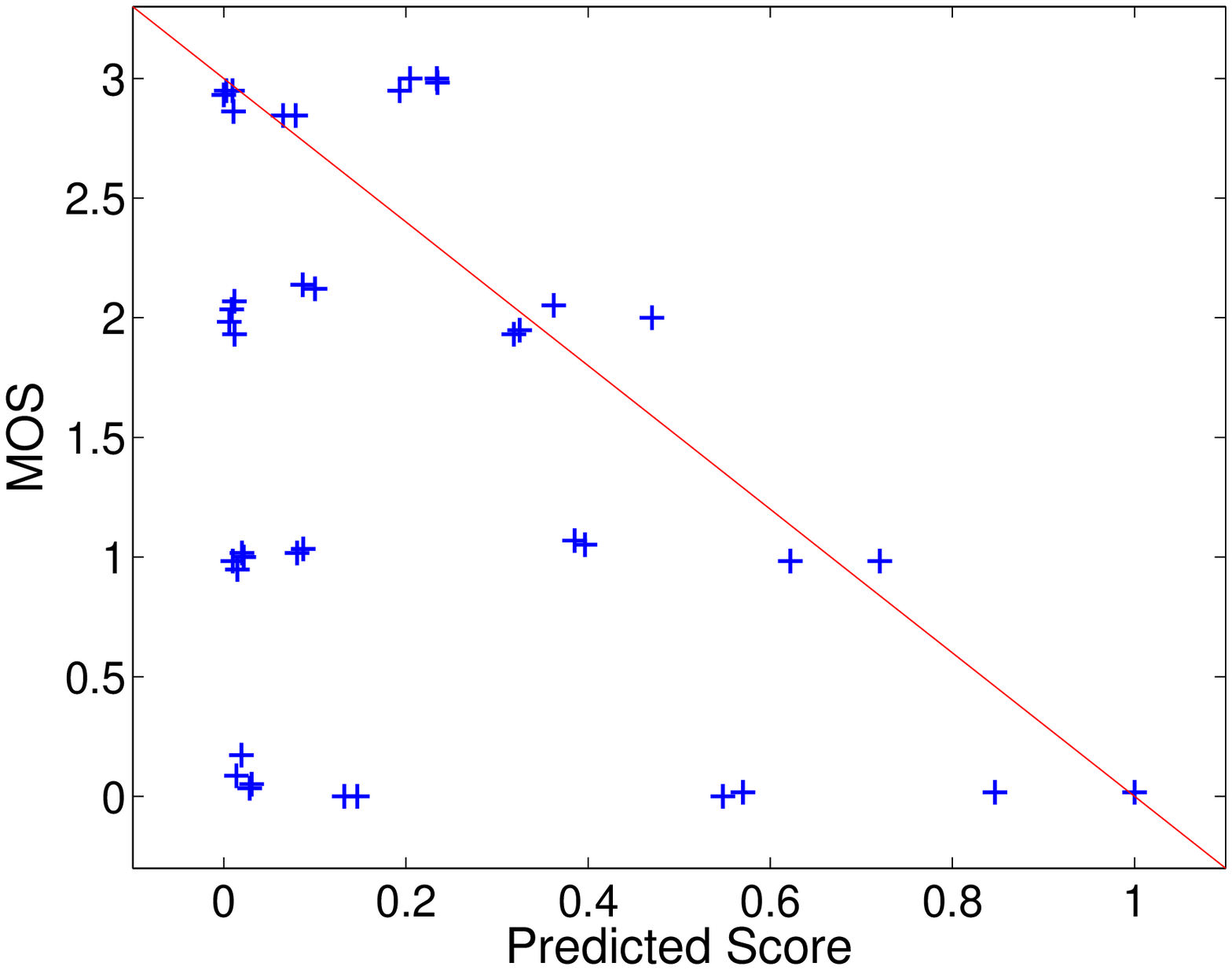}}\hspace{0.01\textwidth}
	\subfloat[VIIDEO\label{fig:viideo}]{\includegraphics[width=0.15\textwidth]{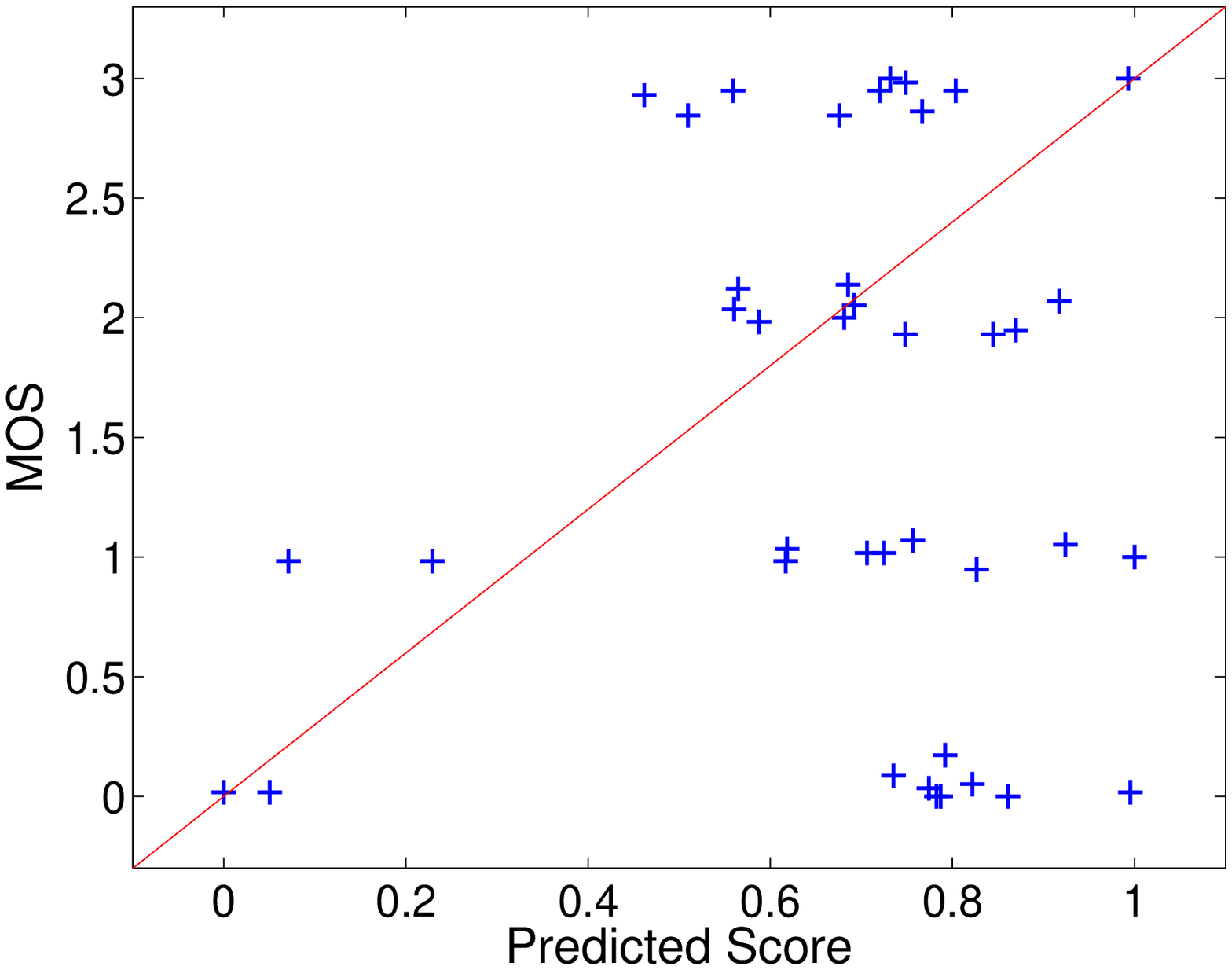}}\hspace{0.01\textwidth}
	\subfloat[V-BLIINDS \label{fig:vbliinds}]{\includegraphics[width=0.15\textwidth]{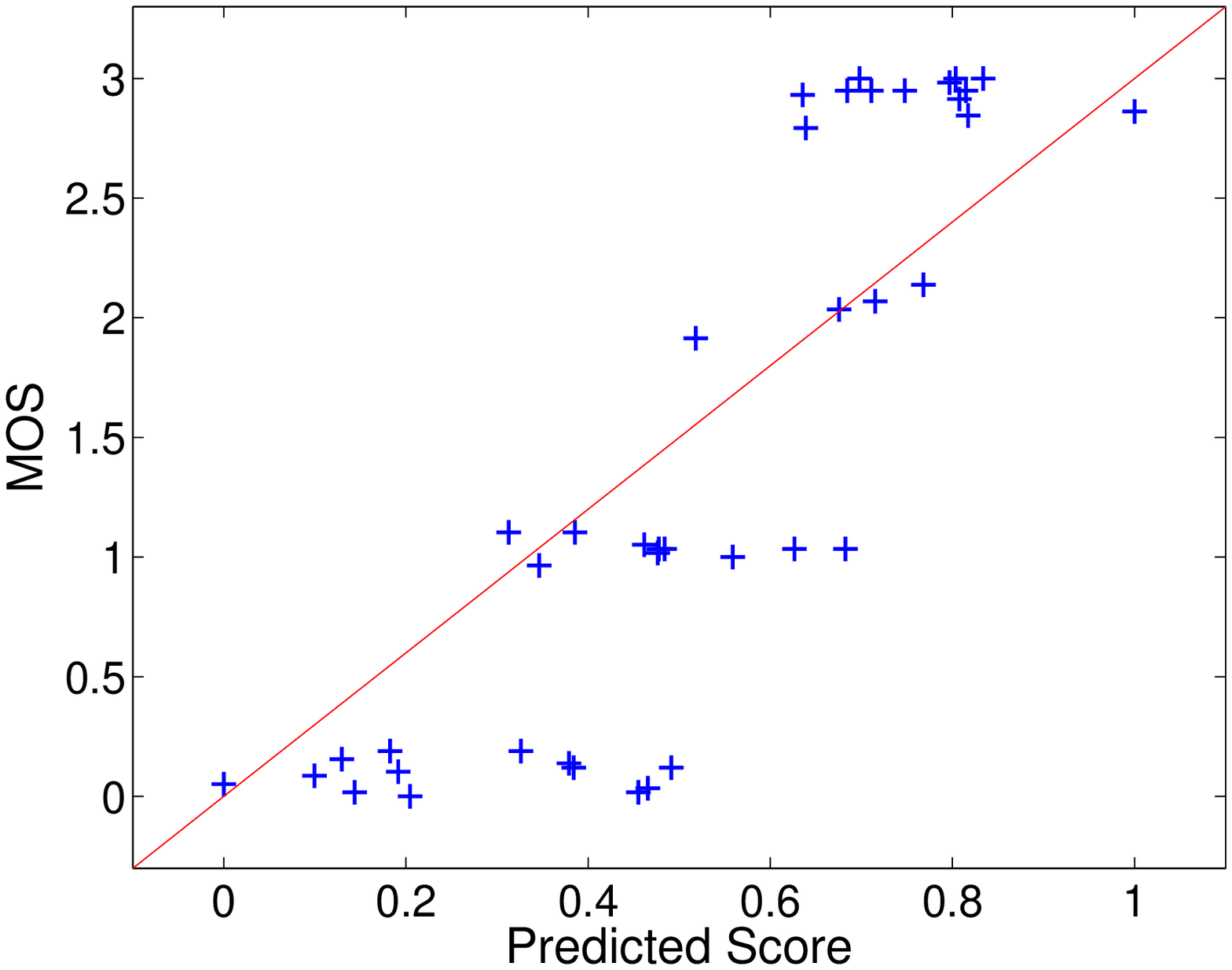}}\hspace{0.01\textwidth}
	\subfloat[Inceptionv3-FT\label{fig:inceptionv3-ft}] {\includegraphics[width=0.15\textwidth]{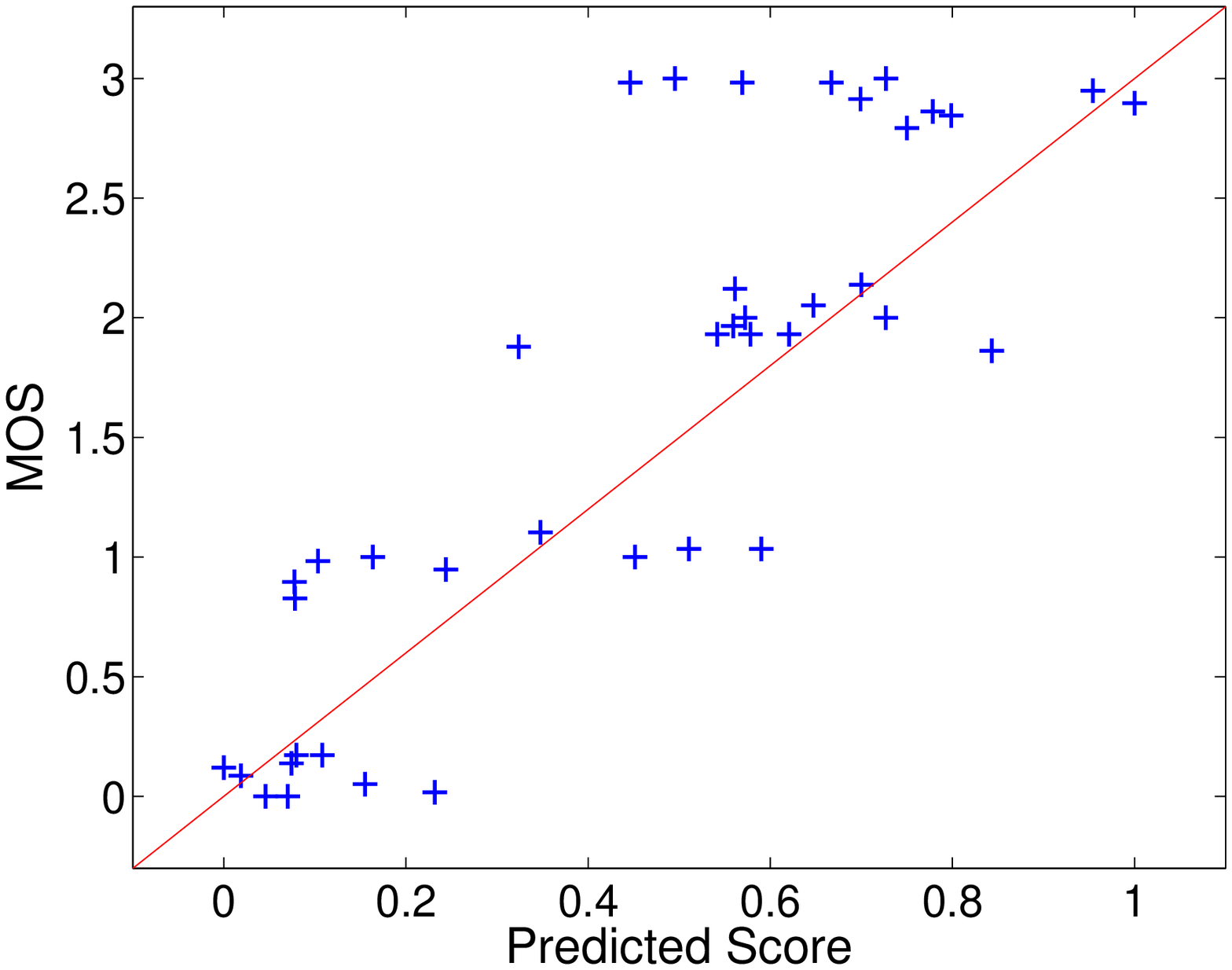}}\hspace{0.01\textwidth}
	\subfloat[VSFA\label{fig:vsfa}] {\includegraphics[width=0.15\textwidth]{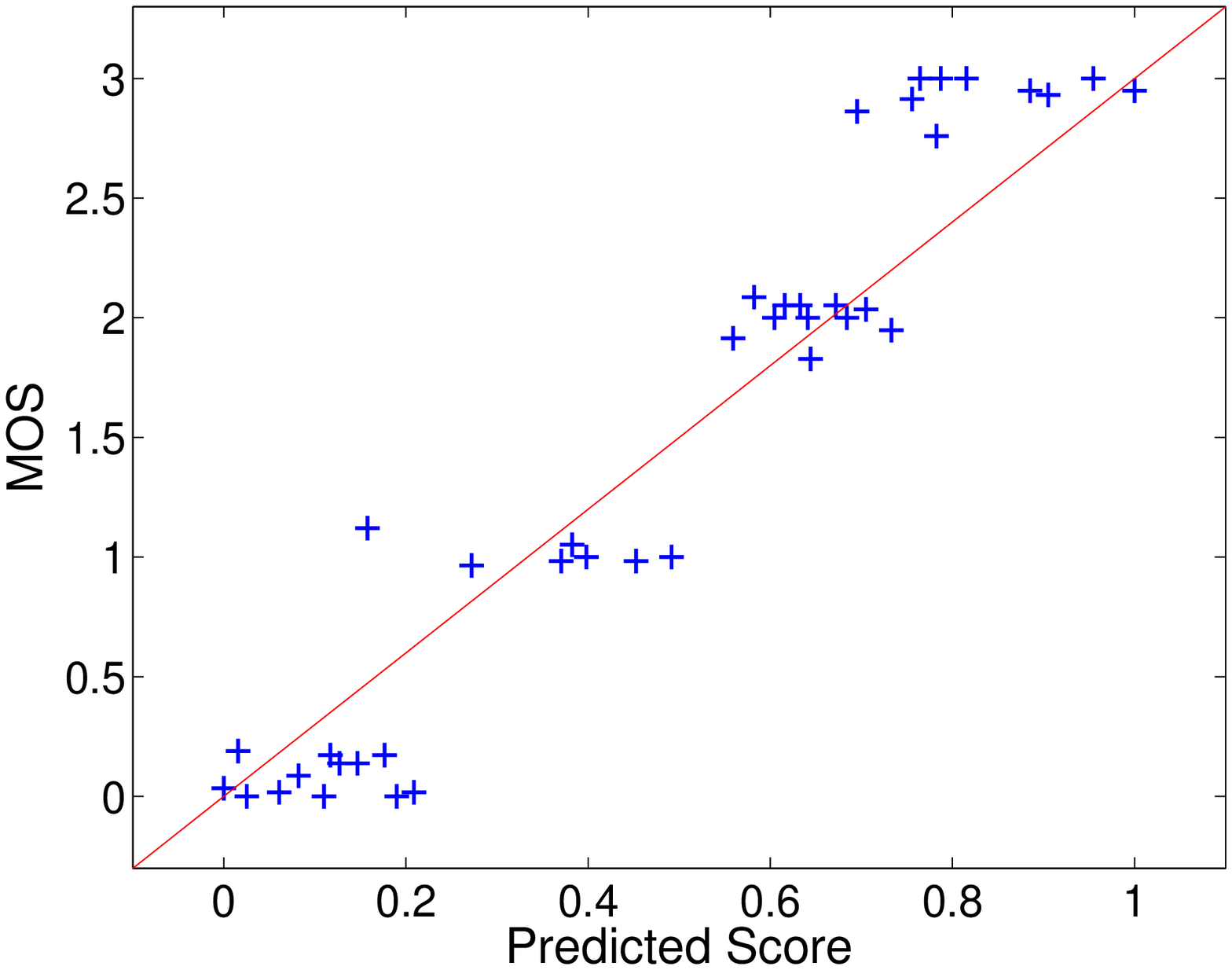}}\hspace{0.01\textwidth}
	\subfloat[TLVQM\label{fig:tlvqm}] {\includegraphics[width=0.15\textwidth]{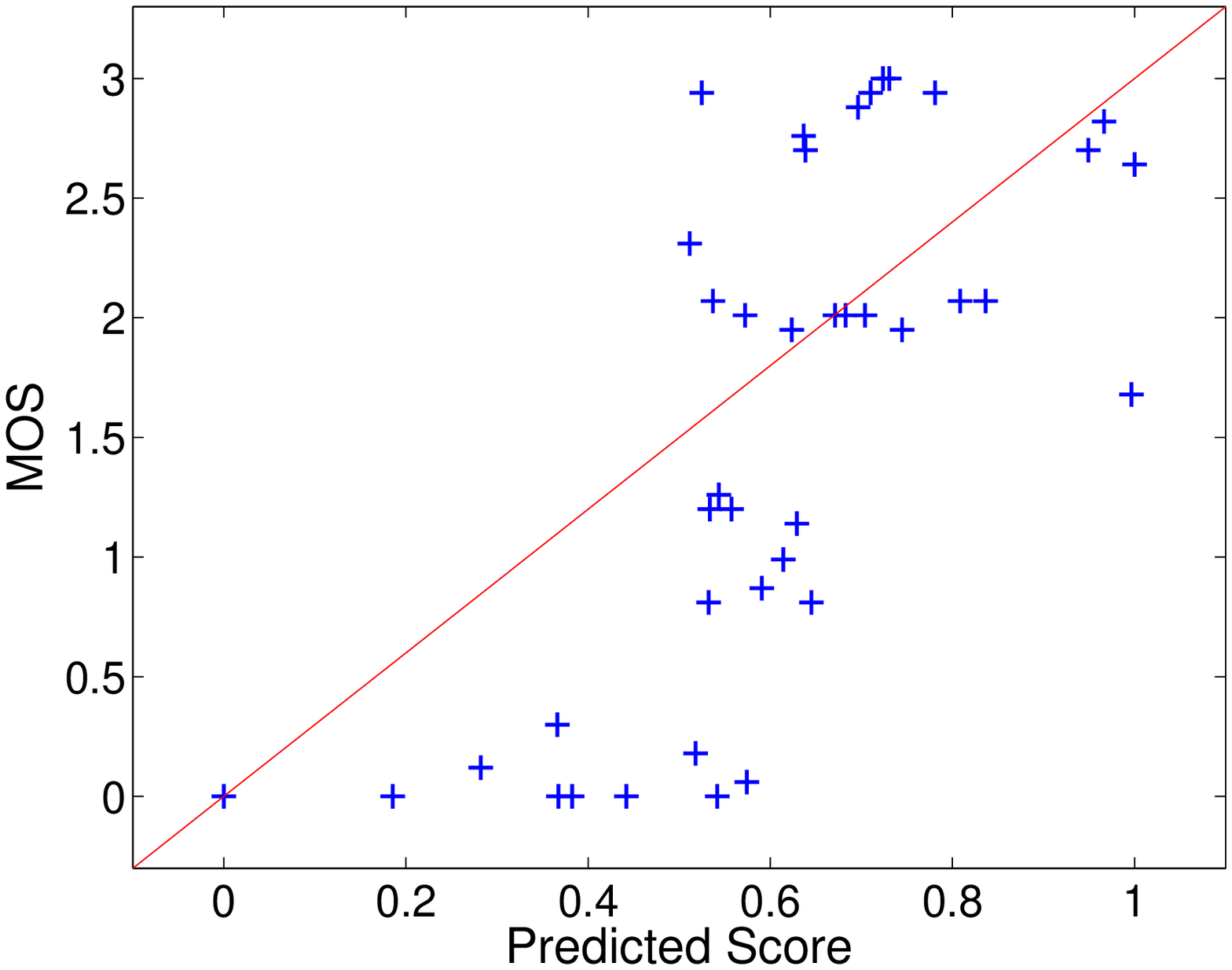}}\hspace{0.01\textwidth}
	\subfloat[CNN-LSTM TLVQM\label{fig:tlvqm_lstm}] {\includegraphics[width=0.15\textwidth]{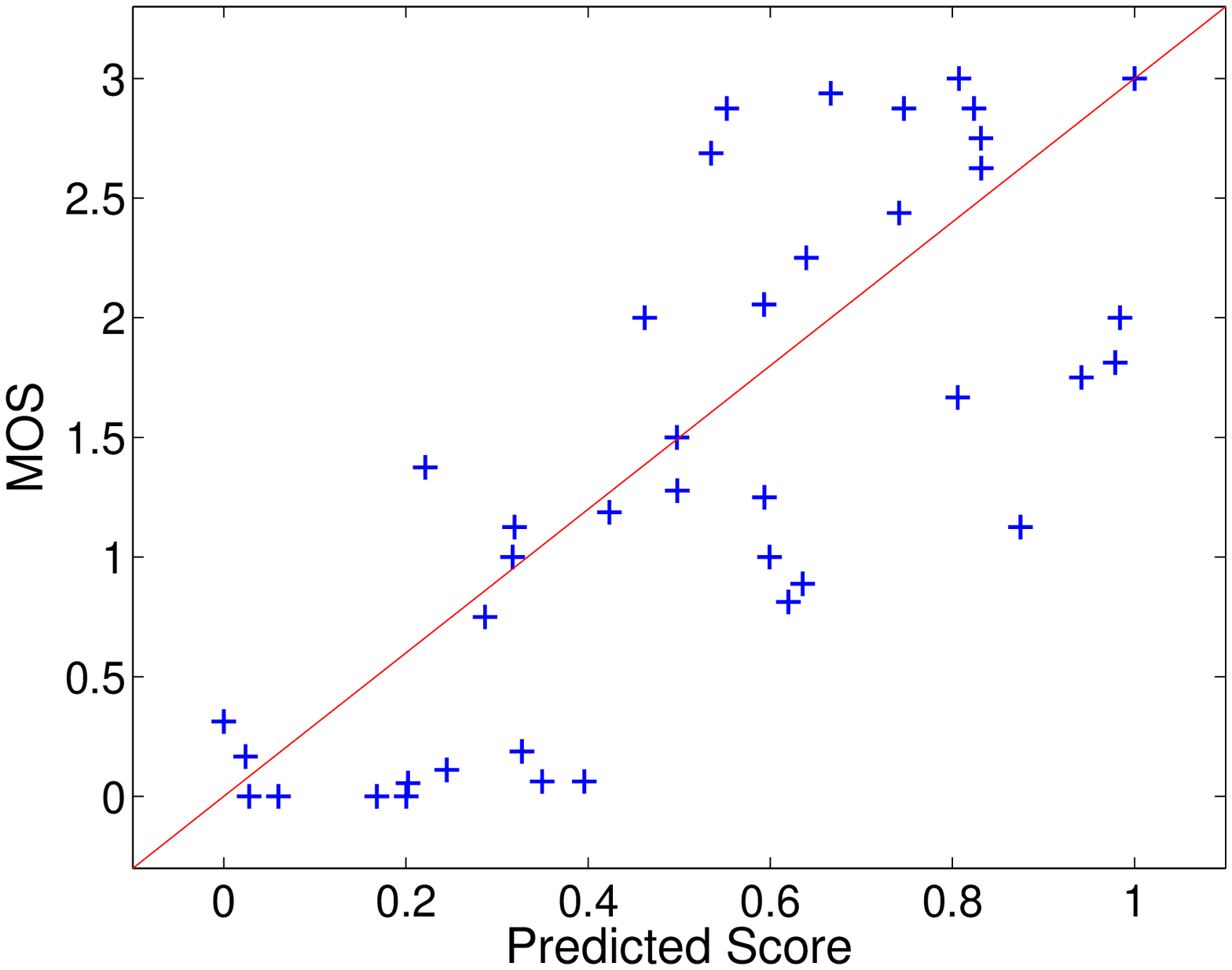}}\hspace{0.01\textwidth}
	\subfloat[CNN-SVR TLVQM \label{fig:tlvqm_svr}] {\includegraphics[width=0.15\textwidth]{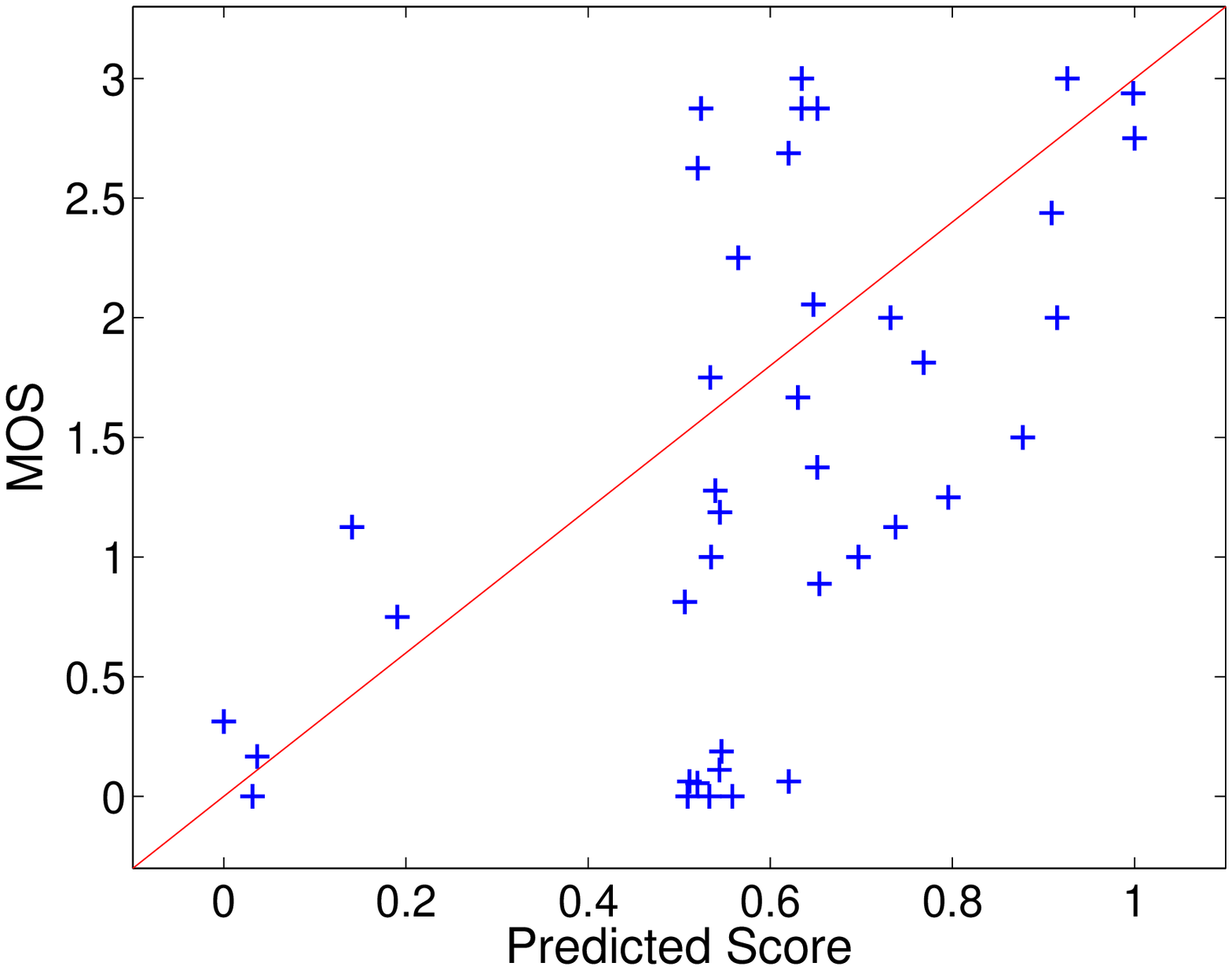}}\hspace{0.01\textwidth}
	\subfloat[Proposed VQP-Net (TL) \label{fig:proposed}]{\includegraphics[width=0.15\textwidth]{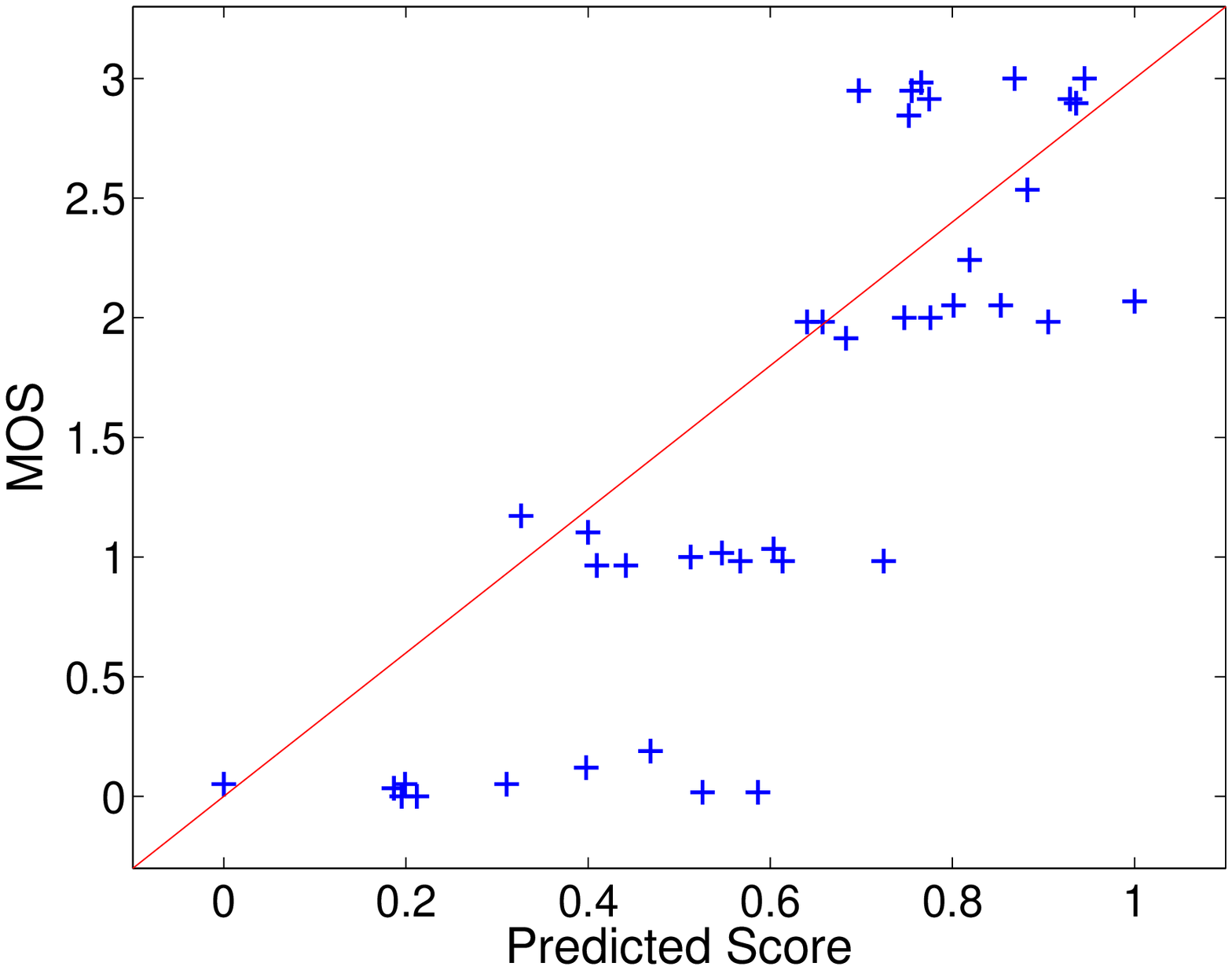}}\hspace{0.01\textwidth}
	\subfloat[Proposed VQP-Net (E2E) \label{fig:proposed_e2e}]{\includegraphics[width=0.15\textwidth]{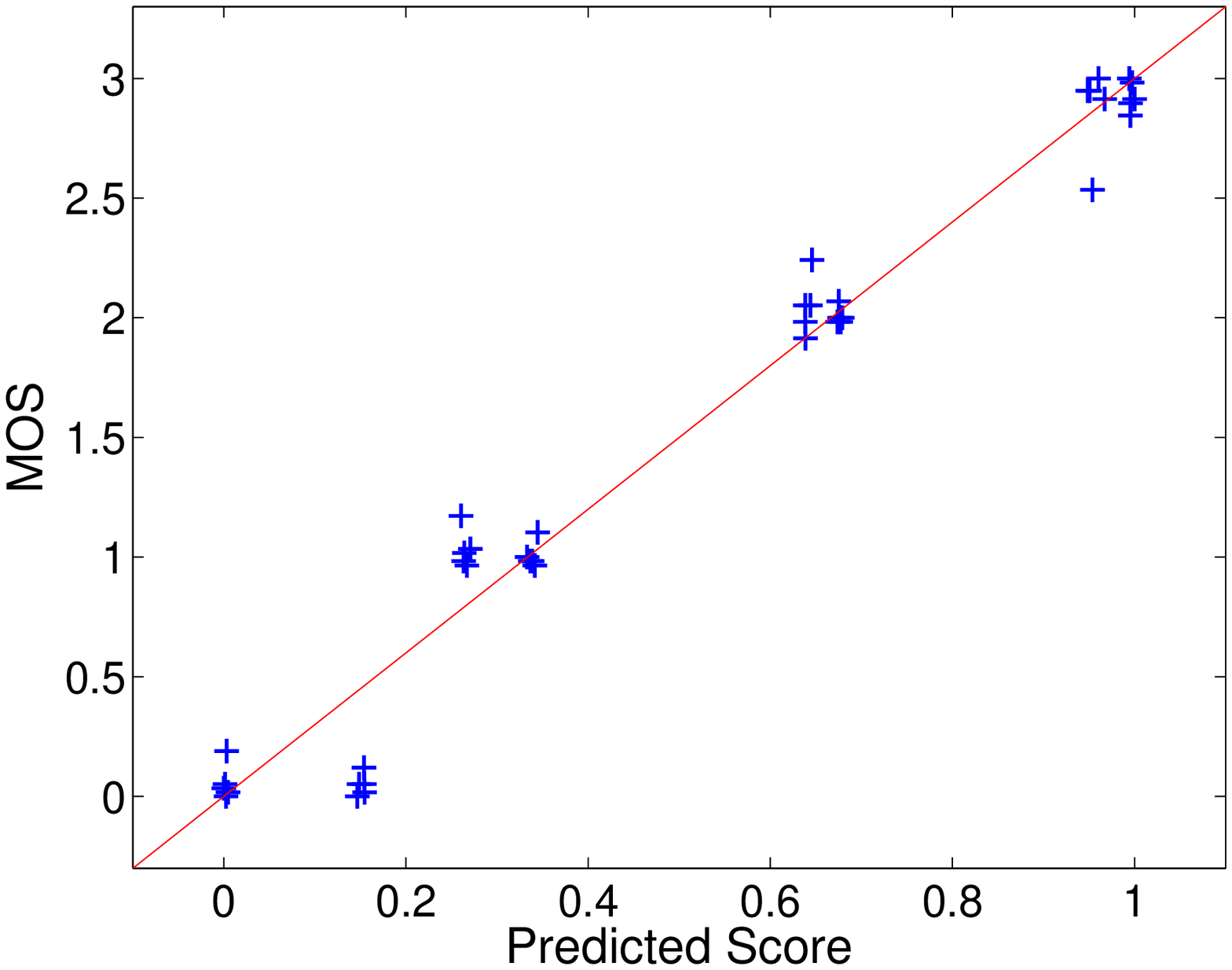}}
	\caption{\label{fig:mos_plots} Subjective (MOS) vs predicted score plots for different VQA metrics.} 
	\vspace{-0.5cm}
\end{figure}

In order to have a better idea of the prediction accuracy for these VQA methods, Figure \ref{fig:mos_plots} shows the scatter plots of subjective scores (MOS) against the normalized predicted scores. All the plots are obtained by using the scores for the test data.  

Ideally, for a good VQA method, the scatter plot should show good linearity, tight clustering and a relatively uniform density along both axes. From the figure, we can see that with IQA metrics and VIIDEO (Figure \ref{fig:psnr} - \ref{fig:viideo}), there is no clear relationship between the two data and they are widely spread throughout. If the points on each plot are closer to the fitted lines, it implies a higher consistency between the subjective and predicted scores. Hence, we can clearly see the performance of the different metrics based on the spread of the data points around the straight line. For instance, for TLVQM based methods (Figure \ref{fig:tlvqm} - \ref{fig:tlvqm_svr}), the data is more spread as compared to the other methods, especially for the CNN-SVR TLVQM metric where the predicted scores are significantly biased towards higher values. On the other hand, for V-BLIINDS and Inceptionv3-FT (Figure \ref{fig:vbliinds} - \ref{fig:inceptionv3-ft}), we can observe that data points are closer and more uniformly distributed as compared to the previously mentioned methods. Similar behavior is also observed with our proposed VQP-Net (TL) method with transfer learning approach. However, with VSFA method (Figure \ref{fig:vsfa}), we see even much better clustering than our proposed transfer learning based approach, even though some of the data points are still away from the fitted line. Undoubtedly, from among all the plots, our proposed VQP-Net (E2E) with the end-to-end learning approach gives the best linearity, clustering and uniform distribution along the two axes as shown in Figure \ref{fig:proposed_e2e}. This further consolidates the observations made based on the correlation coefficient values.

\textbf{Comparison with different temporal pooling approaches}:
In order to show the importance of the additional FCNN model to combine the different frame quality scores, we have performed an ablation study whereby this temporal pooling stage is achieved using other approaches based on neural networks as well as conventional operations. In this respect, we have chosen three different networks which have been considered in previous deep learning based VQA methods  \cite{li2019quality,korhonen2020blind}. These networks are Gated Recurrent Unit (GRU), Long Short-Term Memory (LSTM) and Recurrent Neural Network (RNN). For fair comparison, the overall architecture composed of ResNet followed by the chosen network is trained in an end-to-end manner.
Regarding the conventional temporal pooling methods, we have used arithmetic mean, geometric mean, harmonic mean and median pooling approaches. Note that the latter are simply applied to the quality scores $\tilde{s}^{(m,n)}$ obtained with each frame using the FQP-ResNet model.\\
The results of these different temporal pooling approaches are shown in Table \ref{tab:temp_pooling}. Thus, in case of conventional temporal pooling methods, it can be firstly noticed that both arithmetic mean pooling and median pooling give good results which are close to those obtained using an FCNN with a transfer learning strategy. However, the results of these simple temporal pooling approaches are much less performant than the end-to-end learning version of the FCNN-based approach. Moreover, it can be also observed that the retained FCNN model outperforms significantly the other neural network ones.

\begin{table}[htbp]
	\caption{Comparison of different temporal pooling methods for combining the frame quality scores.}
	\begin{center}
		\resizebox{0.5\textwidth}{!}{\begin{tabular}[width=0.4\textwidth]{|l|c|c|c|}
			\hline
			\rule[-1ex]{0pt}{2.5ex} \textbf{Methods} 
			&  \textbf{PLCC} &  \textbf{SROCC}  &  \textbf{KROCC}  \\
			\hline					
			\rule[-1ex]{0pt}{2.5ex} \textbf{Proposed FCNN (E2E)}&\textbf{0.9899}&\textbf{0.9388} & \textbf{0.7739}\\	
			\hline				
			\rule[-1ex]{0pt}{2.5ex} \textbf{Proposed FCNN (TL)}&0.8992&0.8434 & 0.6494\\	
			\hline			
			\rule[-1ex]{0pt}{2.5ex} \textbf{RNN (E2E)}&0.7768&0.7039&0.5276 \\	
			\hline	
			\rule[-1ex]{0pt}{2.5ex} \textbf{GRU (E2E)}&0.6938&0.6331 & 0.4550 \\	
			\hline
			\rule[-1ex]{0pt}{2.5ex} \textbf{LSTM (E2E)}&0.5434&0.4880 & 0.3461 \\	
			\hline			
			\hline
			\rule[-1ex]{0pt}{2.5ex} \textbf{Arithmetic Mean Pooling}&0.9019  &  0.8459 & 0.6494
			\\	
			\hline	
			\rule[-1ex]{0pt}{2.5ex} \textbf{Geometric Mean Pooling}&0.7728  &  0.7112 & 0.4939
			\\	
			\hline	
			\rule[-1ex]{0pt}{2.5ex} \textbf{Harmonic Mean Pooling}&0.7803  &  0.7190 & 0.5276
			\\	
			\hline	
			\rule[-1ex]{0pt}{2.5ex} \textbf{Median Pooling}&0.9033  &  0.8456 & 0.6520
			\\	
			\hline	
		\end{tabular}}
		\label{tab:temp_pooling}
	\end{center}
	\vspace{-0.7cm}
\end{table}

\section{Conclusion and Perspectives}
\label{sec:conc}
In this work we addressed a rather complex problem that concerns both the identification and classification of distortions and the evaluation of video sequence quality in the context of video-guided surgery. We have then proposed neural networks based approaches for distortion classification and quality prediction in laparoscopic video. While a residual network architecture is employed for the first task, the second one is achieved by adding a fully connected network to combine the quality scores obtained on the different frames of the video sequence. The resulting architecture for the second task is trained using transfer learning and end-to-end learning approaches. The obtained results confirm the benefits of the proposed approaches (i.e. the
proposed end-to-end learning approach) compared to the recent state-of-the-art methods. 
In future work, it would be interesting to enrich the database with other videos of different organs and surgical scenarios. It would be also interesting to investigate new perceptual loss functions for training the overall architecture. 


\bibliographystyle{IEEEtran}
\bibliography{refs}

\end{document}